\documentclass[aps,prl,amsmath,twocolumn,superscriptaddress]{revtex4-2}
\usepackage{amsmath}
\usepackage{amssymb}
\usepackage{graphicx}
\usepackage{tikz}
\usepackage{bm} 
\usepackage{hyperref}

\usepackage{color}

\DeclareMathOperator{\const}{const}

\usepackage{graphicx}
\usepackage{dcolumn}
\usepackage{bm}

\begin{document}

\title{Dynamics of topological defects after a photo-induced melting of a
charge-density wave}

\author{Andrei E. Tarkhov}
\email{atarkhov@bwh.harvard.edu}
\affiliation{Laboratory for the Physics of Complex Quantum Systems, Moscow Institute of Physics and Technology, Dolgoprudny 141700, Russia}
\affiliation{Division of Genetics, Department of Medicine, Brigham and Women’s Hospital and Harvard Medical School, Boston, MA 02115, USA}
\affiliation{Skolkovo Institute of Science and Technology, Moscow 121205, Russia}

\author{A.V. Rozhkov}%
\affiliation{Laboratory for the Physics of Complex Quantum Systems, Moscow Institute of Physics and Technology, Dolgoprudny 141700, Russia}
\affiliation{Institute for Theoretical and Applied Electrodynamics, Russian Academy of Sciences, Moscow 125412, Russia}

\author{Alfred Zong}
\affiliation{Massachusetts Institute of Technology, Department of Physics, Cambridge, MA 02139, USA}
\affiliation{University of California at Berkeley, Department of Chemistry, Berkeley, CA 94720, USA}

\author{Anshul Kogar}
\affiliation{Massachusetts Institute of Technology, Department of Physics, Cambridge, MA 02139, USA}

\author{Nuh Gedik}
\affiliation{Massachusetts Institute of Technology, Department of Physics, Cambridge, MA 02139, USA}

\author{Boris V. Fine}
\email{boris.fine@uni-leipzig.de}
\affiliation{Laboratory for the Physics of Complex Quantum Systems, Moscow Institute of Physics and Technology, Dolgoprudny 141700, Russia}
\affiliation{Skolkovo Institute of Science and Technology, Moscow 121205, Russia}
\affiliation{Institute for Theoretical Physics, University of Leipzig, Leipzig 04103, Germany}%

\date{\today}

\begin{abstract}
Charge-density-wave order in a solid can be temporarily ``melted'' by a strong laser pulse.
Here we use the discrete Gross-Pitaevskii equation on a cubic lattice to simulate the recovery of the CDW long-range phase coherence following such a pulse. Our simulations indicate that the recovery process  is dramatically slowed down by the three-dimensional topological defects -- CDW dislocations -- created as a result of strongly nonequilibrium heating and cooling of the system. Overall, the simulated CDW recovery was found to be remarkably reminiscent of a recent pump-probe experiment in LaTe$_3$.


\end{abstract}

\maketitle




Topological defects in the form of vortices play
a prominent role in nonequilibrium and equilibrium statistical
physics of ordered
systems~\cite{berezinsky1970destruction, berezinsky1972destruction,
kosterlitz1972long,svistunov1991highly, kagan1992erratum, kagan1994kinetics,
berloff2002scenario,yoshino2002extended, proctor2014random, garanin2015ordered,
kobayashi2016quench, kumar2017ordering, vasin2019description}. 
In particular, vortices are known to emerge through the Kibble-Zurek
mechanism~\cite{kibble1976topology, kibble1980some, zurek1985cosmological,
zurek1996cosmological},
when the system is cooled across a phase transition. 
In the Kibble-Zurek setting, the system is initially at equilibrium, and the cooling process is supposed to be adiabatic everywhere except for a close vicinity of the phase transition temperature.
It is still an outstanding question: What happens after a quench across a phase transition, where the initial state is far from equilibrium, and the energy flow from the system is so fast that  the system does not reach quasi-equilibrium at any time during the quench?    



In the present work, we address the above question in the context of the
recent pump-probe experiment of
Ref.~\cite{zong2019evidence}
on the light-induced melting of a charge-density wave (CDW) in
$\text{LaTe}_3$.
The experiment monitored the recovery of the CDW order after its
destruction by a laser pulse. The experimental system  was
initially in the low-temperature ordered phase; then the part of the system
responsible for the CDW order was rapidly heated up and cooled down, so
that it did not have time to reach the high-temperature equilibrium. It was
observed~\cite{zong2019evidence} 
that the amplitude and the phase of the CDW responded
qualitatively differently: the amplitude recovered quickly, whereas, for
sufficiently high pump pulse intensity, the recovery of phase coherence
took much longer~\cite{dolgirev2020amplitude}.
It was conjectured in
Ref.~\cite{zong2019evidence}
on the basis of indirect experimental evidence that the observed slowdown
of the CDW-order recovery is due to the slow dynamics of the photoinduced
topological defects -- CDW dislocations. Topological defects within the CDW phase were  also invoked in other contexts in Refs.~\cite{domain_walls_trigo2021exper,wandel2020lightenhanced,
late3_OPcompet2020kogar_exper,defect2019review_mihalovic,vogelgesang2018phase, duan2021optical,
photoind_tas2_2020exper,photo_induced_zong2021review}.

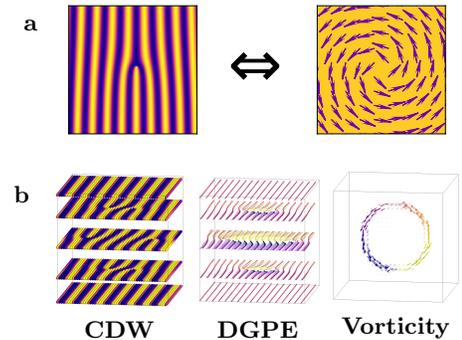
\begin{figure}[ht!]
    \begin{tikzpicture}
    \node[inner sep=0pt] (duck) at (0,0)
    {\includegraphics[width=0.6\columnwidth]{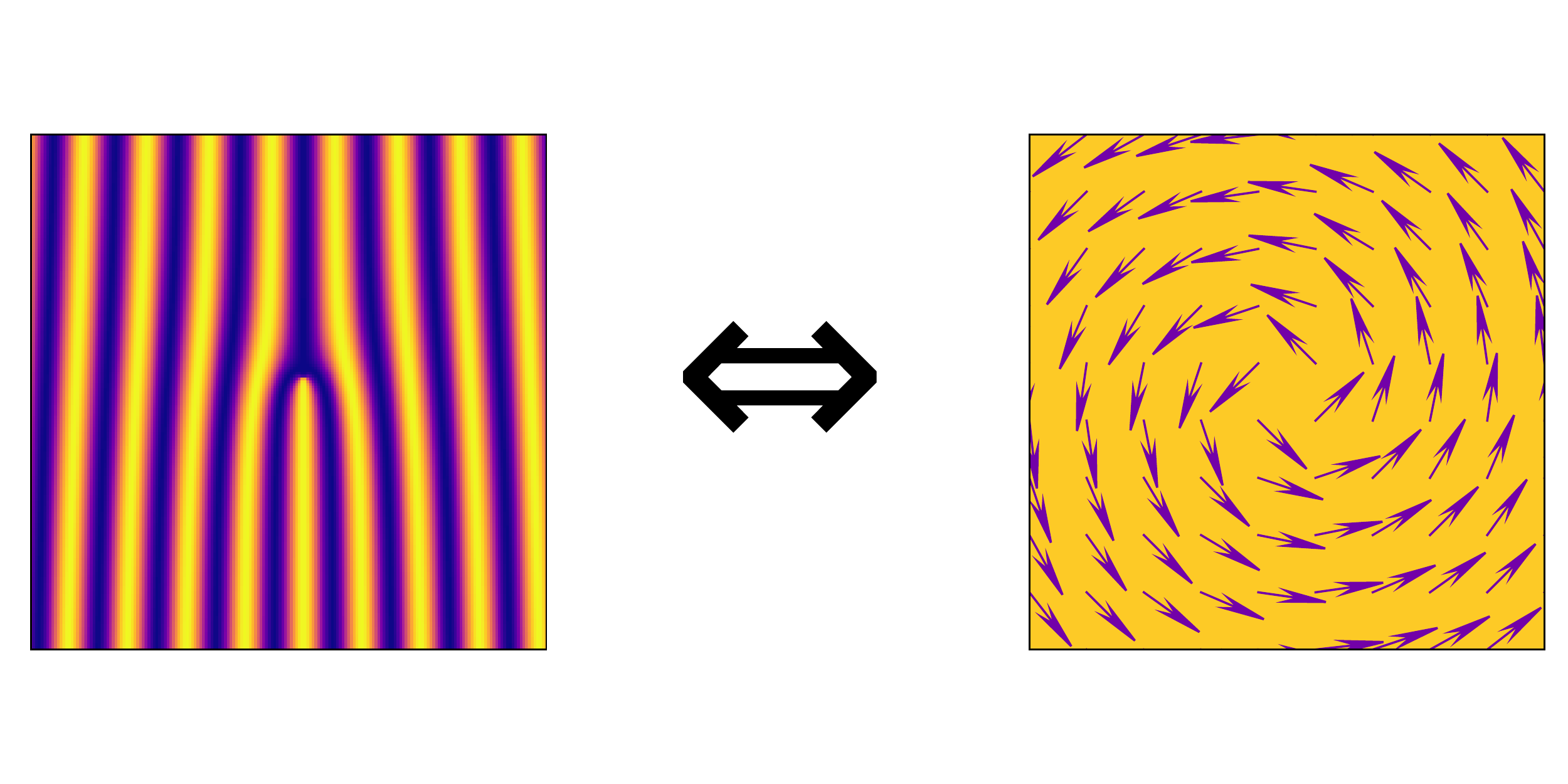}};
    \node[align=center,fill=white] at (-3.,0.6) {\textbf{a}};
    \end{tikzpicture}
    
    \begin{tikzpicture}
    \node[inner sep=0pt] (duck) at (0,0)
    {\includegraphics[width=0.2\columnwidth]{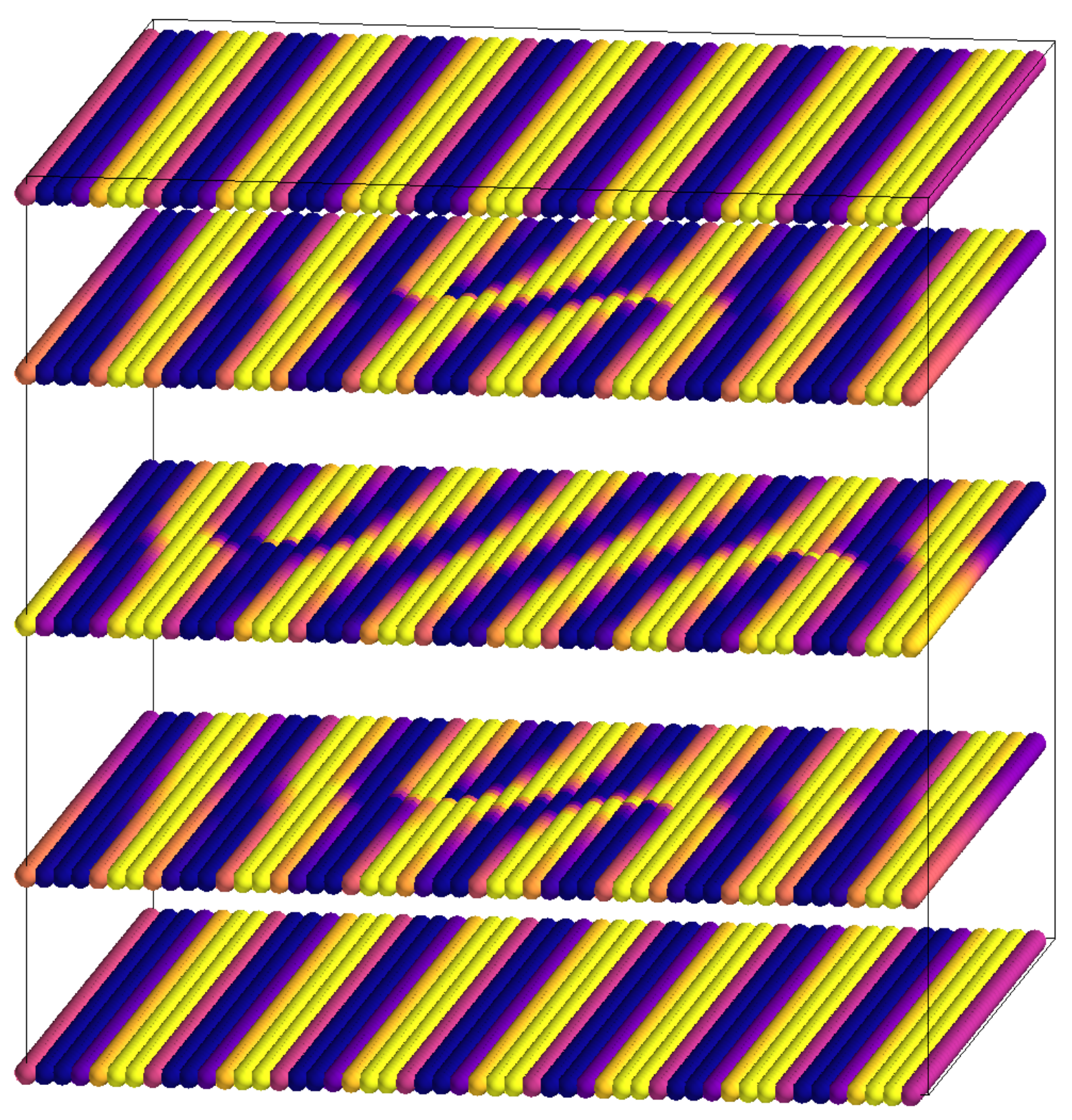}};
    \node[align=center,fill=white] at (-1.3, 0.6) {\textbf{b}};
    \node[align=center,fill=white] at (-0.0, -1.2) {\textbf{CDW}};
    \end{tikzpicture}
    \begin{tikzpicture}
    \node[inner sep=0pt] (duck) at (0,0)
    {\includegraphics[width=0.2\columnwidth]{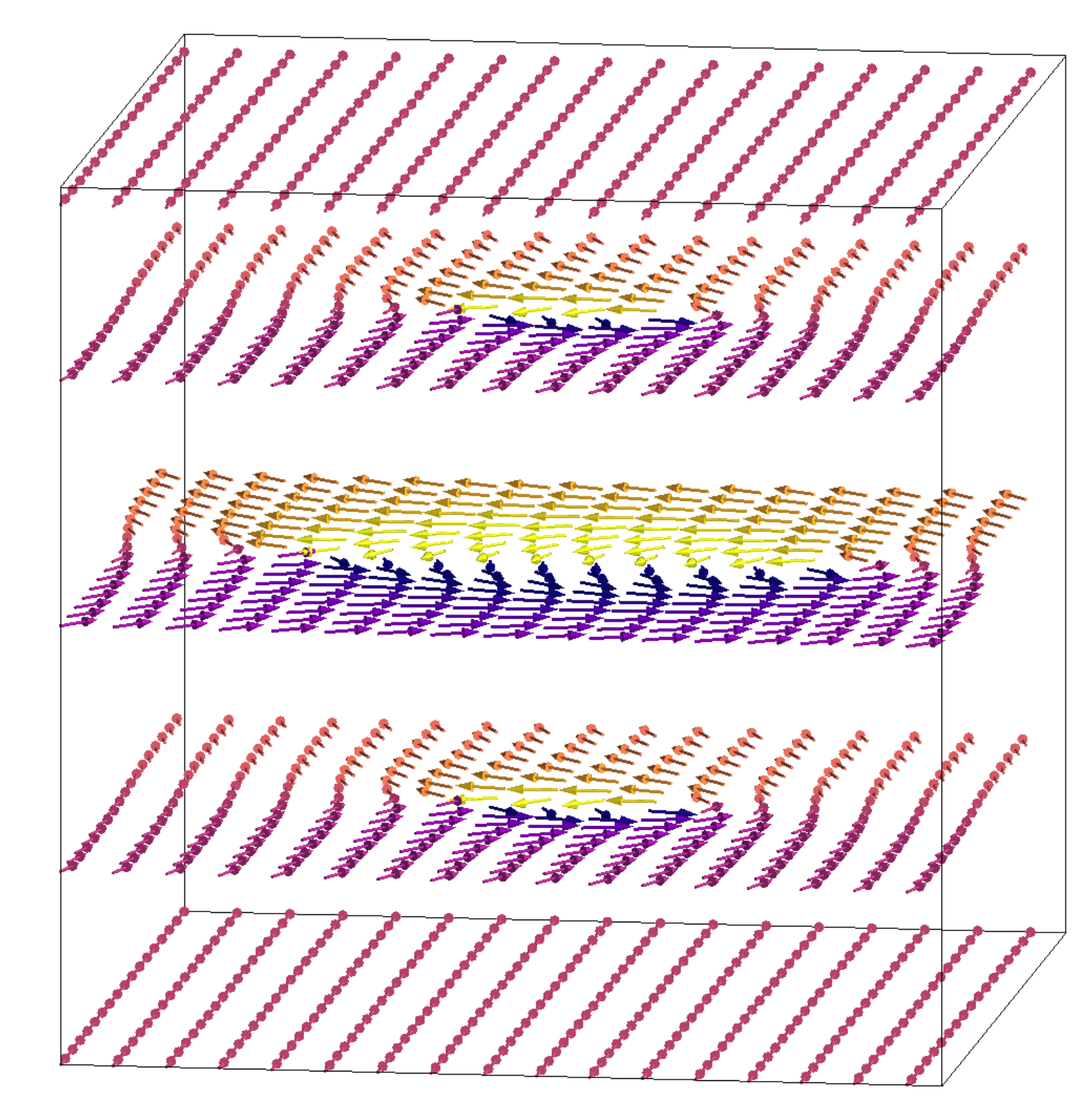}};
    \node[align=center,fill=white] at (-0.0, -1.2) {\textbf{DGPE}};
    \end{tikzpicture}
    \begin{tikzpicture}
    \node[inner sep=0pt] (duck) at (0,0)
    {\includegraphics[width=0.2\columnwidth]{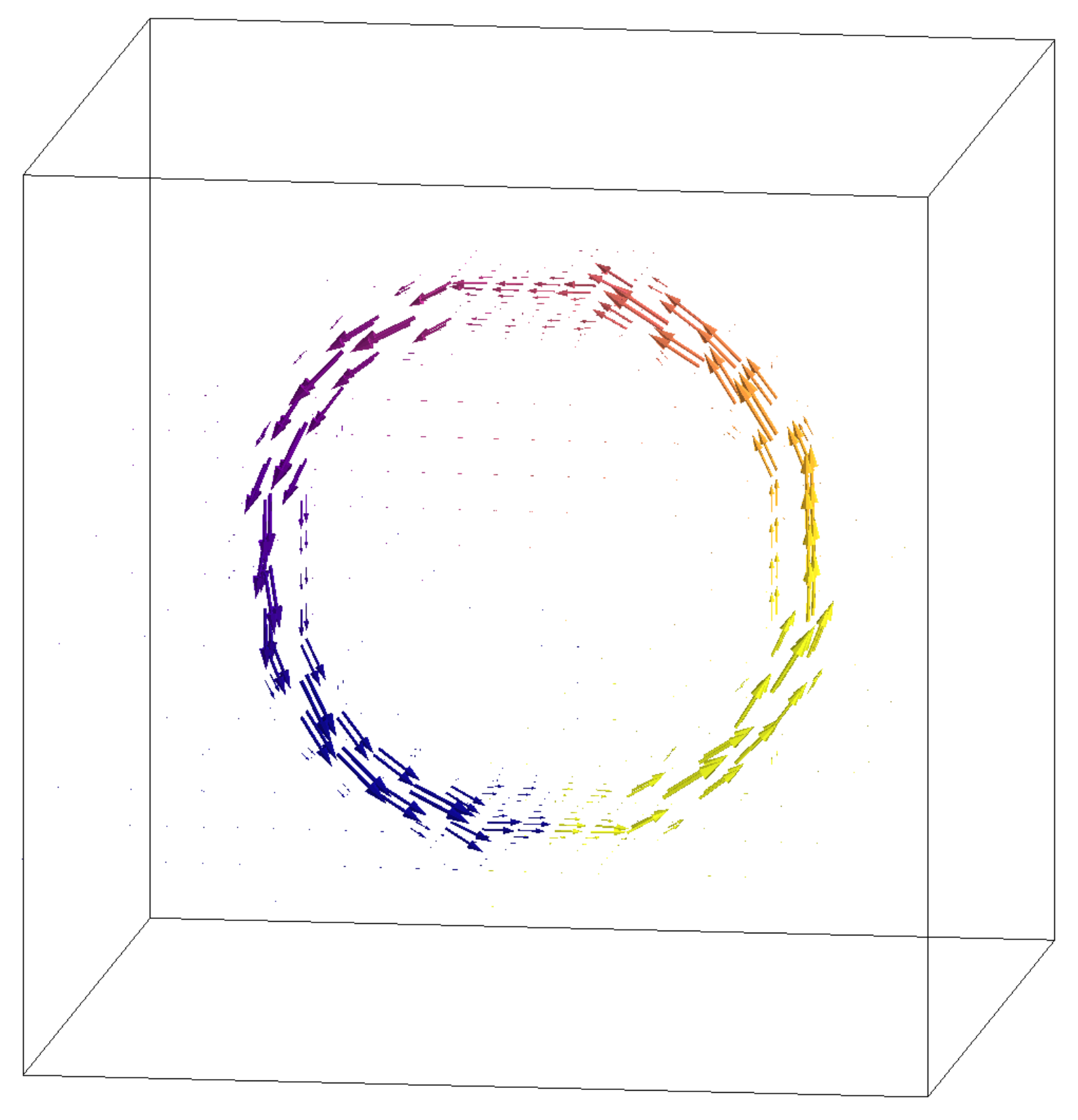}};
    \node[align=center,fill=white] at (-0.0, -1.2) {\textbf{Vorticity}};
    \end{tikzpicture}
\caption{
Sketch of the mapping of a CDW dislocation $u \cos\left[\mathbf{q}_{\rm CDW} \mathbf{r}_m + \phi(\mathbf{r}_m)\right] $ onto a vortex of the DGPE $U(1)$
lattice variables $u \exp\left[ i \phi(\mathbf{r}_m)\right]$. (a) 2D slice of the CDW dislocation  (left) vs. 2D slice of the DGPE vortex (right).  
(b) Three-dimensional representations of a vortex loop: CDW (left) vs. DGPE lattice (middle) vs. vorticity of the DGPE lattice defined in the text (right).
\label{fig:cdw2dgpe}
}
\end{figure}

The translational symmetry breaking associated with the onset of a CDW can
be equivalently viewed as the breaking of the $U(1)$ symmetry with respect
to the phase of the CDW order parameter. In such a picture, the CDW
dislocations are mapped onto vortices in the underlying $U(1)$-ordered
state  (see
Fig.~\ref{fig:cdw2dgpe}).


In the present work, we simulate the principal features of the CDW phase dynamics with the help of
the discrete Gross-Pitaevskii equation (DGPE), which also exhibits a $U(1)$ phase transition. We
implement the heating-cooling quench, perform the numerical imaging of vortices, trace their
dynamics and, thereby demonstrate that they are responsible for the slowdown of the order parameter
recovery. The  basic character of our simulations suggest that the observed non-equilibrium
phenomenology is applicable to a broad class of physical systems. 

{\it General considerations. ---}
CDW is a modulation of electronic charge density accompanied by a periodic
modulation of ionic positions
\begin{equation}
\label{quench:eq:cdw_lattice_modulation}
\mathbf{r}_m \to \mathbf{r}_m +
	u \, \mathbf{e} \cos\left(\mathbf{q}_{\rm CDW} \mathbf{r}_m + \phi\right),
\end{equation}
where 
$\mathbf{r}_m$
is the position of the
$m$'th ion,
$\mathbf{q}_{\rm CDW}$
the CDW wave vector,  $u$ and $\phi$ the amplitude and the phase of the
CDW, and $\mathbf{e}$ the unit vector along the CDW lattice displacement
direction~\cite{dolgirev2020amplitude}.
One can choose the complex number
$u\exp(i\phi)$
as the CDW order parameter. By allowing this order parameter to vary slowly
in space and time, 
$u(\mathbf{r}, t)\exp\left(i\phi(\mathbf{r}, t)\right)$,
one can then describe the low-frequency dynamics of a CDW.

\begin{figure}[ht!]
    \begin{tikzpicture}
    \node[inner sep=0pt] (duck) at (0,0)
    {\includegraphics[width=0.9\columnwidth]{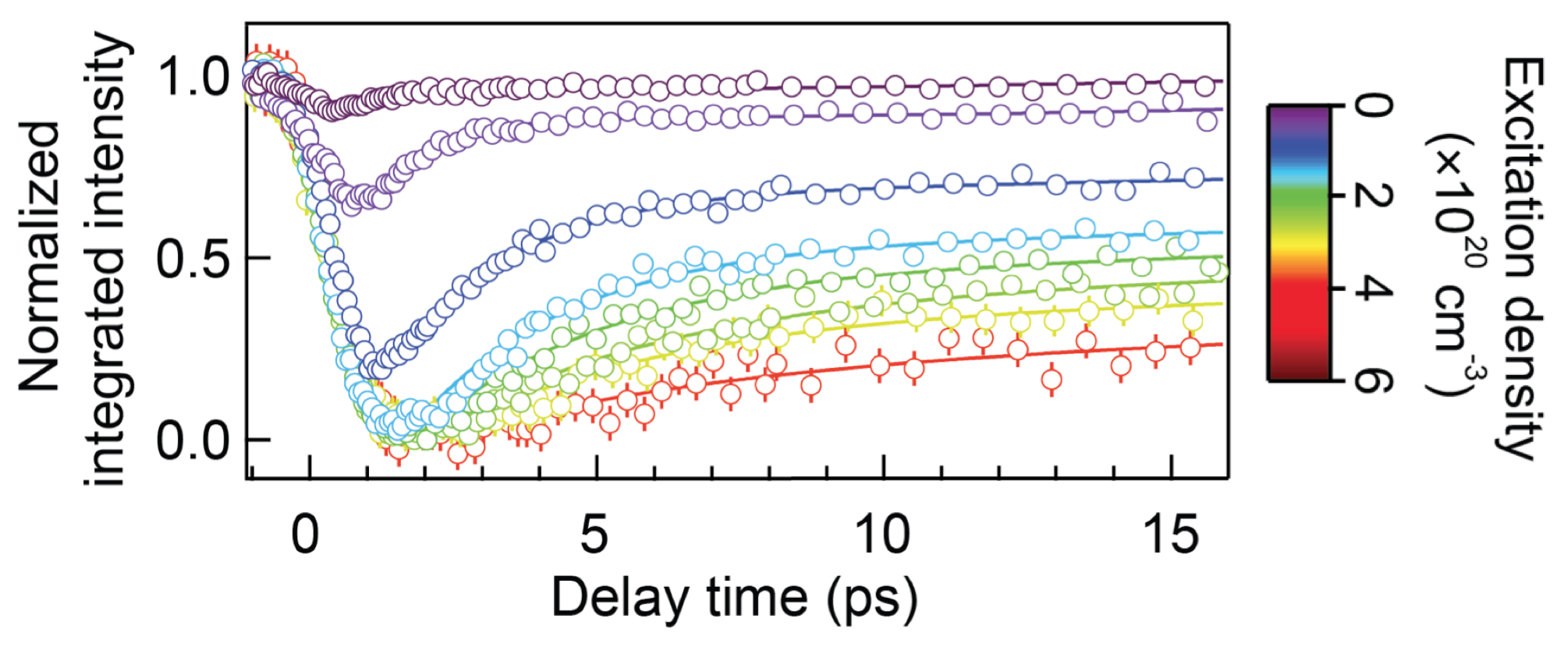}};
    \node[align=center,fill=white] at (-3.8, 1.8) {\textbf{a}};
    \end{tikzpicture}
    ~
    \begin{tikzpicture}
    \node[inner sep=0pt] (duck) at (0,0)
    {\includegraphics[width=0.85\columnwidth]{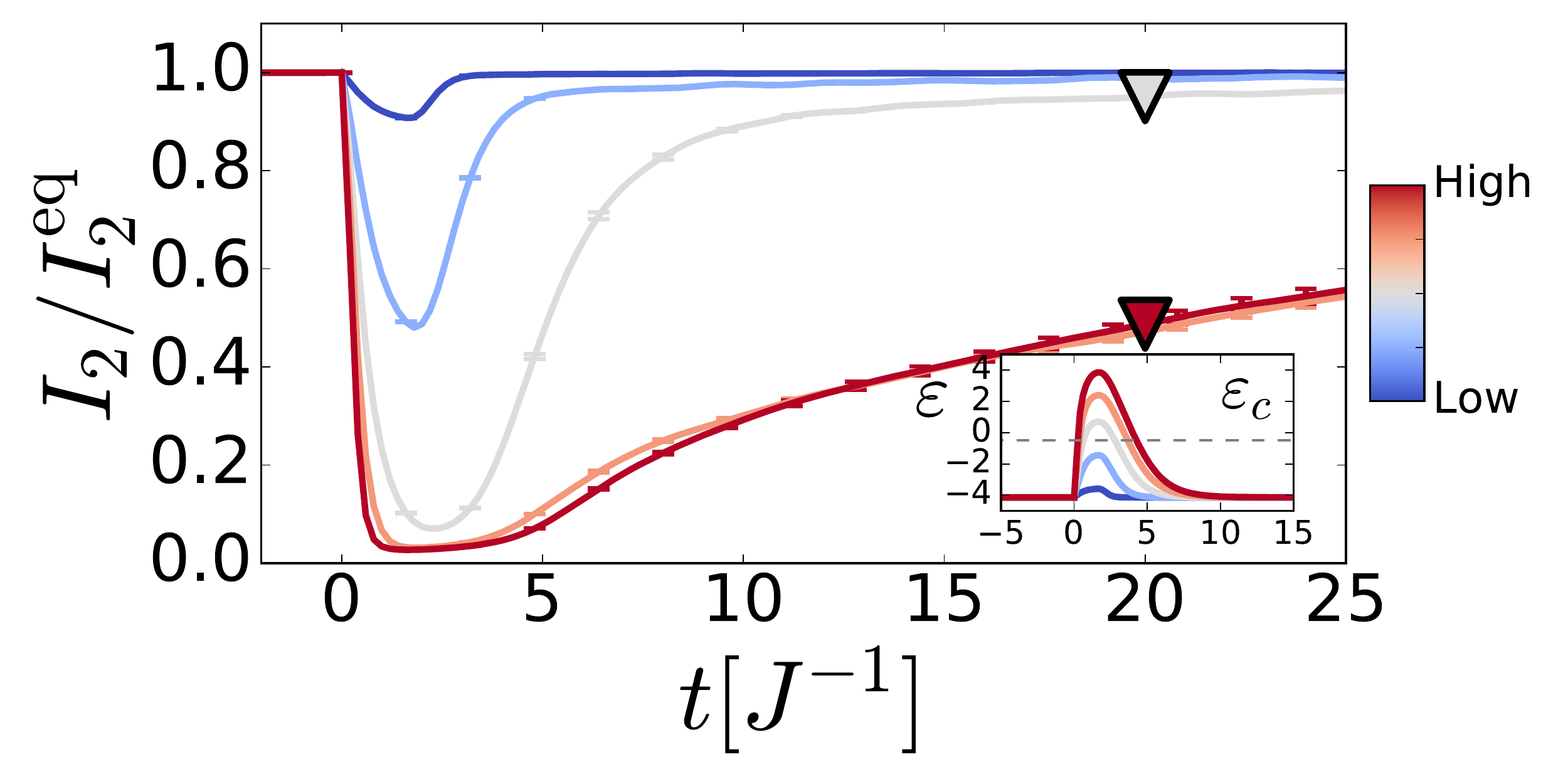}};
    \node[align=center,fill=white] at (-3.8, 2.0) {\textbf{b}};
    \end{tikzpicture}
\caption{(a)~Time evolution of experimentally measured 2D-integrated intensity of the
CDW electron diffraction peak after photo-excitation, for several laser
photo-excitation densities (from
Ref.~\cite{zong2019evidence}).
(b)~Numerically-calculated 2D-integrated intensity
$I_2 \equiv \sum_{k_x, k_y} |\psi (k_x, k_y, 0)|^2$ tracking the ordering in the system [see
Eq.~(\ref{eq::I2_def})]
as a function of time for different quench intensities.
The color of the curves encodes different quench intensities as indicated on the
colorbar. Each numerical data point represents the average of $40$ quench realizations, and the error bars represent the statistical uncertainty associated with this average.
Inset: Variation of the energy density during quenches. Horizontal line
marks critical energy density $\varepsilon_c$, see
Fig.~\ref{quench:fig:thermodynamics}(a), above which the order
is destroyed.
\label{quench:fig:experiment_CDW}
}
\end{figure}

In
Fig.~\ref{quench:fig:experiment_CDW}(a),
we reproduce one of the experimental results of
Ref.~\cite{zong2019evidence},
namely, the time dependence of the 2D-integrated intensity of the CDW diffraction peak for
different fluences of the laser pump pulse. 
This intensity is proportional to the squared amplitude of the CDW order parameter.
For smaller pulse fluences, the
order parameter initially becomes smaller but never vanishes entirely
and then quickly recovers to almost the initial value. For higher fluences,
the CDW first disappears, i.e. becomes totally melted, but then it recovers
over much longer times, in fact, not reaching the equilibrium value on the
timescale accessible in the experiment. The subsequent theoretical analysis
of 
Ref.~\cite{dolgirev2020amplitude}
concluded that the above slow recovery cannot be explained by the amplitude
dynamics of the CDW order (see
also~\cite{photo_meltCDW2021exper, zong_CDW2019slowing_down_exper}).
The amplitude takes about $2$~ps to relax~\cite{zong2019evidence,dolgirev2020amplitude}
It is the phase relaxation happening afterwards that we aim at
describing in this paper.

{ \it The model. ---} Our numerical model, DGPE, is a
space-discretized version of the continuous Gross-Pitaevskii equation
(GPE). Both can be considered as possible dynamical extensions of the
static XY-model. The GPE describes the low-temperature dynamics of
superfluid Bose gases, while the DGPE is often used to describe bosons on
optical lattices.
We note that both the spatially inhomogeneous CDW order
$u(\mathbf{r})\exp\left(i\phi(\mathbf{r})\right)$
and the DGPE  share the
$U(1)$
structure of the dynamical variable, and both also exhibit  a spontaneous symmetry breaking transition into
a $U(1)$-ordered state.
The actual dynamics of the inhomogeneous CDW is different and, in many
respects, more complex than that of DGPE. 
Real CDWs are dissipatively coupled with the electrons and with the
lattice degrees of freedom, while the DGPE dynamics is time-reversible. Yet, due to the large number of degrees of freedom involved in the simulations, DGPE delivers an effectively dissipative environment for the smaller number of variables associated with the formation of topological defects over
larger spatial scales. It can be reasonably expected that the detailed
character of the thermal bath is not
crucial here, and, therefore, the thermal bath of the real CDW
system can be efficiently replaced by the bath associated with the
microcanonical thermalizing dynamics of the DGPE lattice.

To connect the CDW dynamics and the DGPE, we divide the crystal lattice
into cells of sizes
$l_x \times l_y \times l_z$,
where
$l_x$,
$l_y$,
and
$l_z$
are  of the order of the CDW coherence lengths for the respective
directions. 
The mapping is achieved by
associating the DGPE lattice variable
$\psi_j$
with the average of the CDW order parameter over the CDW cell around position
$\mathbf{r}_j$:
\begin{equation}
\label{quench:eq:CDW2DGPE_mapping_order_parameter}
\psi_j \sim \left<u(\mathbf{r}) e^{i\phi(\mathbf{r})}\right>_{j\text{th cell}}.
\end{equation}
Given the crystal anisotropy of LaTe$_3$, the coherence length for the direction perpendicular to the Te layers,  $l_z$, should be significantly smaller than $l_x$ and $l_y$. We expect that $l_x$, $l_y$ and $l_z$  can be chosen such that the renormalized phase stiffness  between adjacent {\it anisotropic} unit cells can be modelled with the help of the same constant $J$~\cite{cardy1996scaling,carlson1999classical,supplemental} for all three directions. 
Each cell is thus represented by a single site of the cubic DGPE
lattice. The modelling is classical, since, for large enough cells, the
quantum fluctuations can be neglected. 

\begin{figure}[ht!]
    \centering
    \includegraphics[width=1.\columnwidth]{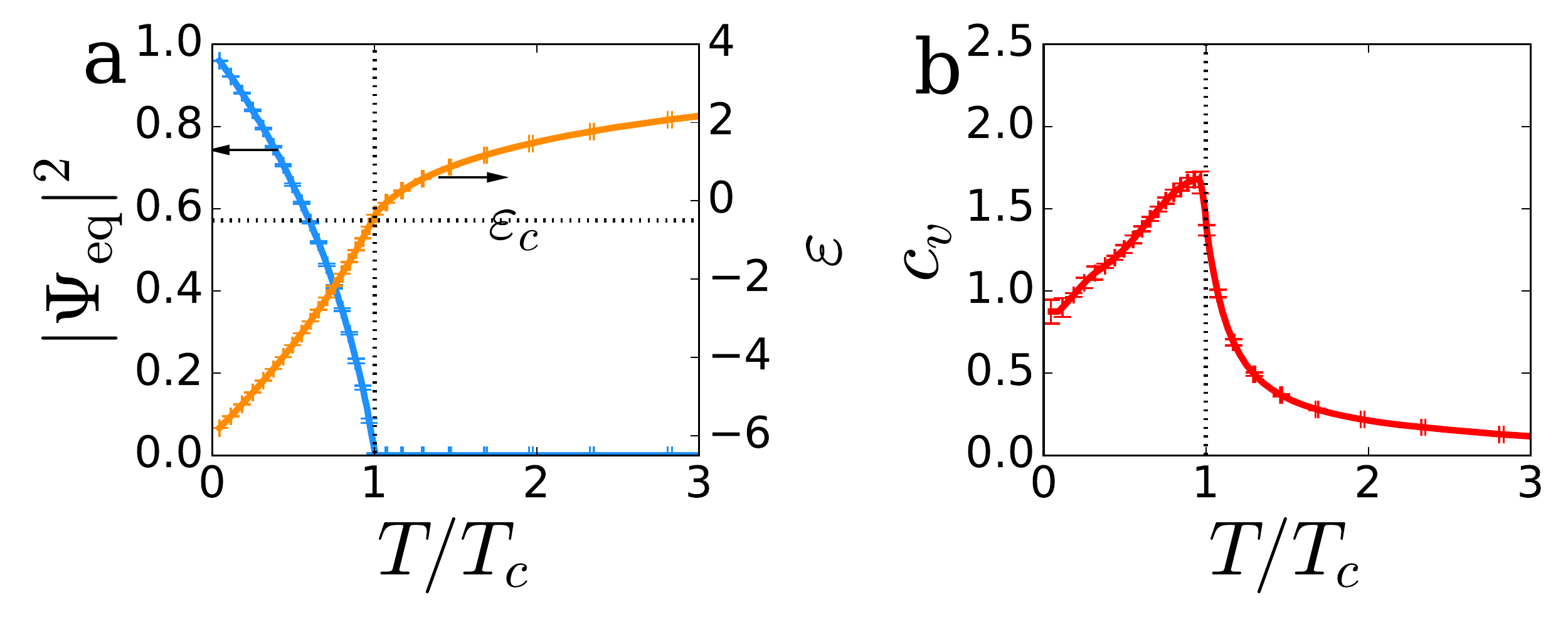}
\caption{Temperature dependence of various parameters for the 3D DGPE lattice: (a) the equilibrium order parameter $|\Psi_{\rm{eq}}|^2$ and the energy density $\varepsilon$, (b) specific heat $c_v$.
\label{quench:fig:thermodynamics}
}
\end{figure}

The time evolution of
$\psi_j$
is to be modelled by the DGPE on a cubic lattice:
\begin{equation}
\label{Intro:eq:equations-motion-DGPE}
i\frac{d\psi_{j}}{dt}
=
-J\sum_{k\in \text{NN}(j)}\psi_{k}+U\left|\psi_{j}\right|^{2}\psi_{j},
\end{equation}
where the summation over $k$
runs through all nearest-neighbors
$\text{NN}(j)$
of site $j$, coefficient $J$ is the ``hopping'' parameter and $U$ is
the interaction parameter. The lattice has $V$ sites. In the simulations, the typical lattice dimensions were $50 \times 50 \times 50$. DGPE conserves the total energy of the lattice
$E \equiv \sum_{j}
	\left(-J\sum_{k\in \text{NN}(j)}
		\psi_{k}\psi^*_{j}+\frac{U}{2}\left|\psi_{j}\right|^{4}\right)$
as well as the ``total norm''
$N \equiv \sum_{j} |\psi_{j}|^2$. We also define the energy density
$\varepsilon \equiv E/V - U/2$ and the ``norm density'' $n \equiv N/V$. The character of the DGPE solutions is determined by the dimensionless parameter $g \equiv U n/J$, which, for the reasons explained later, was chosen to be equal to 10. 
Below, except for the period of quench, we simulate an isolated DGPE
lattice, which, when perturbed, is observed to exhibit dynamic thermalization to a
microcanonical temperature $T$ that was determined numerically as in
Ref.~\cite{de2015chaotic,tarkhov2020ergodization}.
The computed temperature is a monotonic function of 
$\varepsilon$. For $g=10$, $J=1$, $n=1$, the DGPE exhibits a
spontaneous symmetry breaking transition at the critical temperature
$T_{\rm c} \approx 4.25$
into low-temperature ordered state characterised by the
$U(1)$
order parameter 
$\Psi (t) = |\Psi (t) | e^{i \phi(t)} \equiv \frac{1}{\sqrt{n}V} \sum_j \psi_j (t)$.
In equilibrium, 
$|\Psi| = |\Psi_{\rm eq}| (T)$,
where
$|\Psi_{\rm eq}| (T)$
is a decreasing function of $T$ for
$T< T_{\rm c}$,
vanishing above
$T_{\rm c}$.
Both 
$\varepsilon (T)$
and
$|\Psi_{\rm eq}|^2 (T)$
are plotted in
Fig.~\ref{quench:fig:thermodynamics}(a).
The heat capacity 
$c_v (T) \equiv d \varepsilon / dT $ is shown
in
Fig.~\ref{quench:fig:thermodynamics}(b); it exhibits a very recognizable cusp at 
$T=T_{\rm c}$ corresponding to $\varepsilon=\varepsilon_c$.

In principle, Eq.~(\ref{Intro:eq:equations-motion-DGPE}) can be generalized to include disorder representing the CDW pinning centers. However, in the target experiment~\cite{zong2019evidence}, the slowdown of the recovery of the CDW coherence was observed on the spatial scale where no evidence of the presence of the pinning centers was detected. 

{ \it Parameters. ---} To finalize the mapping, we provide the
order-of-magnitude estimates for the three parameters of the cubic DGPE lattice, namely, $J$, 
$n$,
and $U$.

Since the CDW coherent domains are supposed to be  larger than the crystal
unit cell, the CDW phase dynamics should occur at typical frequencies smaller
than the Debye frequency
$\Omega_{\rm D}$.
Hence, the paramater $J$, which determines the characteristic frequency
of the DGPE dynamics, can be roughly estimated as
\begin{equation}
J = \Omega_{\rm D} \frac{a}{l},
    \label{J}
\end{equation}
where
\mbox{$a \equiv (a_x a_y a_z)^{1/3}$} and \mbox{$l \equiv (l_x l_y l_z)^{1/3}$}  are the geometrical averages
of the crystal lattice periods
$a_x, a_y, a_z$ and
the CDW coherence lengths $l_x, l_y, l_z$, respectively. 
We use the mean-field theory~\cite{gruner2018density}, which is supposed to be rather crude for LaTe$_3$, to make the order-of-magnitude estimate
$l = \hbar v_{\rm F}/( 1.76 \pi k_{\rm B} T_{\rm c})$,
where
$v_{\rm F} = 2 \varepsilon_{\rm F}/p_{\rm F} $
is the Fermi velocity, and
$T_{\rm c}$
is the CDW phase transition temperature.  Thus
$\varepsilon_{\rm F}  \approx 1.5$~eV~\cite{yao2006theory},
$p_{\rm F} \approx \frac{3\pi}{8}\frac{\hbar}{a}$,
$T_c \sim 700$~K.
As a result, we get
$l \sim 6 a$.
Inserting this together with
$\Omega_{\rm D} \sim 4$\,THz
into Eq.(\ref{J}) gives
$J \sim 4 \text{ ps}^{-1}$.

The typical value for $n$ can be derived from the fact that, at the critical
temperature, the hopping energy is of order of the thermal energy
$J n \sim k_{\rm B} T_{\rm c}$.
Thus:
\begin{equation}
    n \sim \frac{k_{\rm B} T_{\rm c}}{J} \sim \hbar \frac{T_{\rm c}}{T_{\rm D}} \frac{l}{a}.
\end{equation}
At the last step,  we replaced
$\hbar \Omega_{\rm D}$ in Eq.(\ref{J})
with
$k_B T_{\rm D}$,
where
$T_{\rm D} \sim 200 \rm{K}$
is the Debye temperature for LaTe$_3$~\cite{ru2006thermodynamic}. Thus
$n/\hbar \sim 20$,
which is consistent with the classical character of our modelling.

Parameter $U$ can be determined by matching the speed of phasons,
$c_{\rm ph}$
with the speed of low-frequency excitations of the DGPE lattice:
$c_{\rm ph} = l \sqrt{2 J U n}$~\cite{supplemental}.
In turn, $c_{\rm ph}$
can be crudely estimated as the speed of sound~\cite{manley2018supersonic}. It is further shown in~\cite{supplemental} that, under these assumptions, $U \sim 10 \frac{J}{n}$. Therefore, our simulations were done with $g = U n /J = 10$.

\begin{figure}[ht!]
    \centering
    \begin{tikzpicture}
    \node[inner sep=0pt] (duck) at (0,0)
    {\includegraphics[width=0.45\columnwidth]{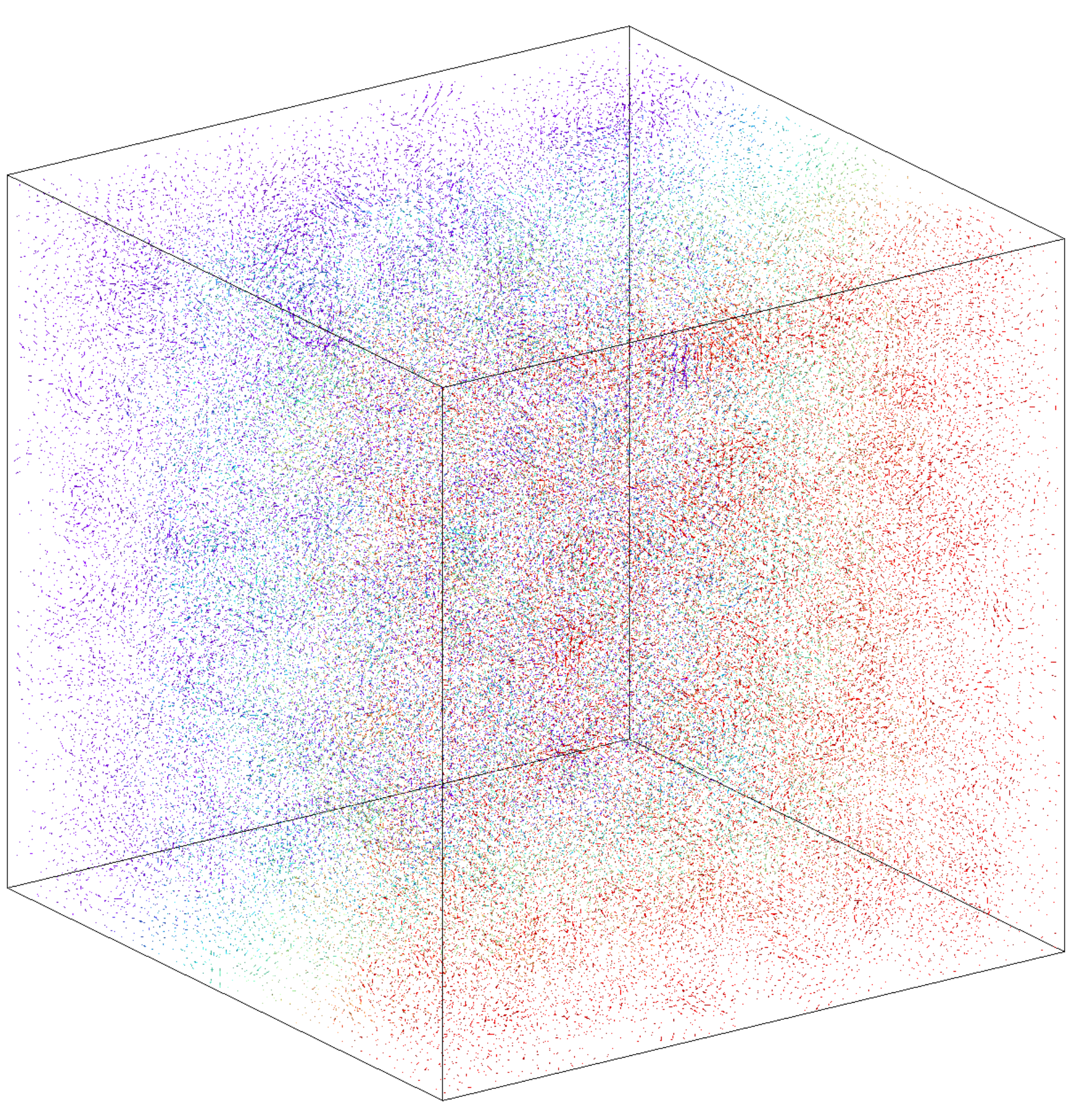}};
    \node[align=center,fill=white] at (-1.5, 1.7) {\textbf{a}};
    \end{tikzpicture}
    \begin{tikzpicture}
    \node[inner sep=0pt] (duck2) at (0,0)
    {\includegraphics[width=0.45\columnwidth]{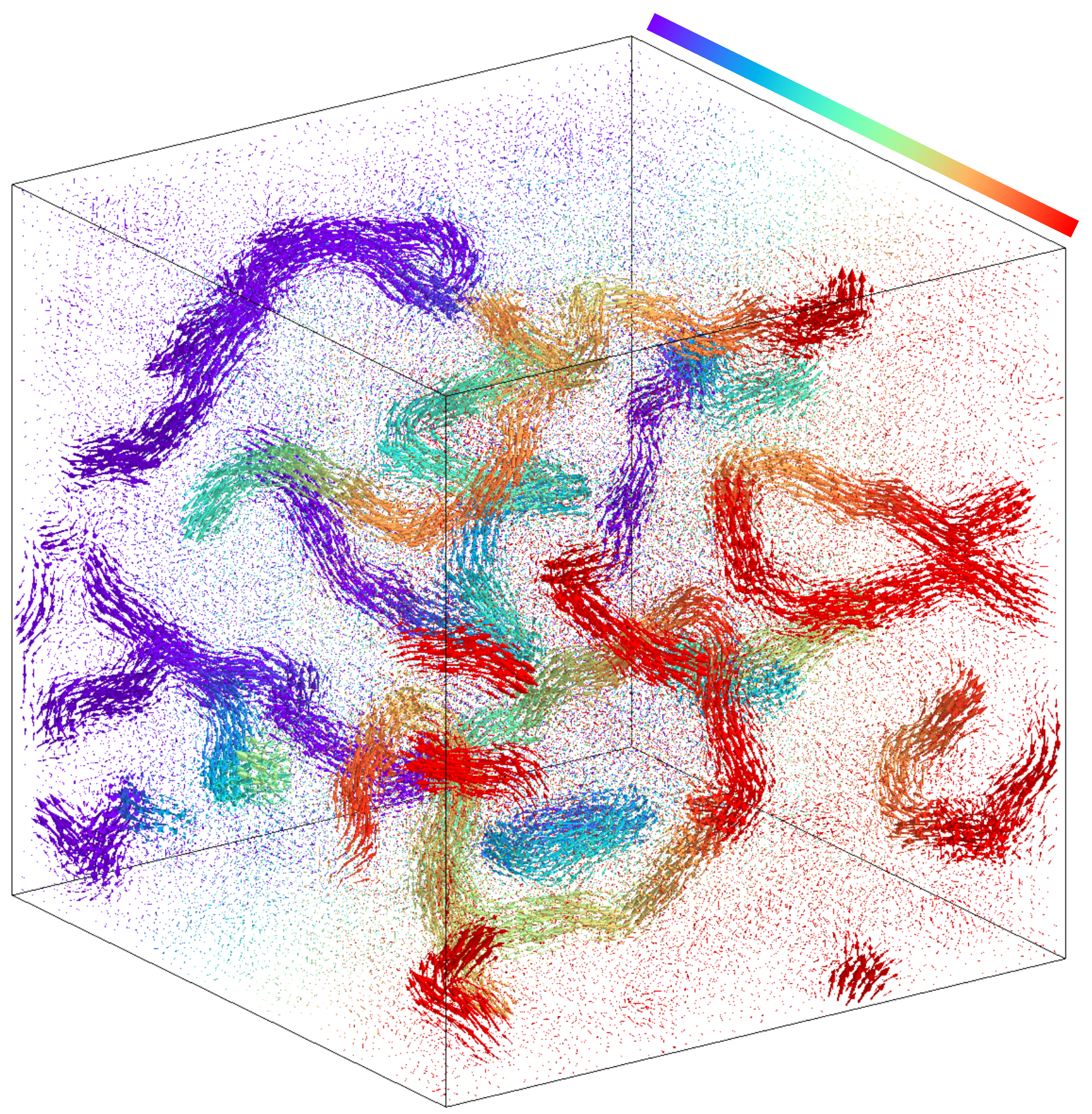}};
    \node[align=center,fill=white] at (-1.5, 1.7) {\textbf{b}};
    \end{tikzpicture}
    
    \begin{tikzpicture}
    \node[inner sep=0pt] (duck3) at (0,0)
    {\includegraphics[width=0.8\columnwidth]{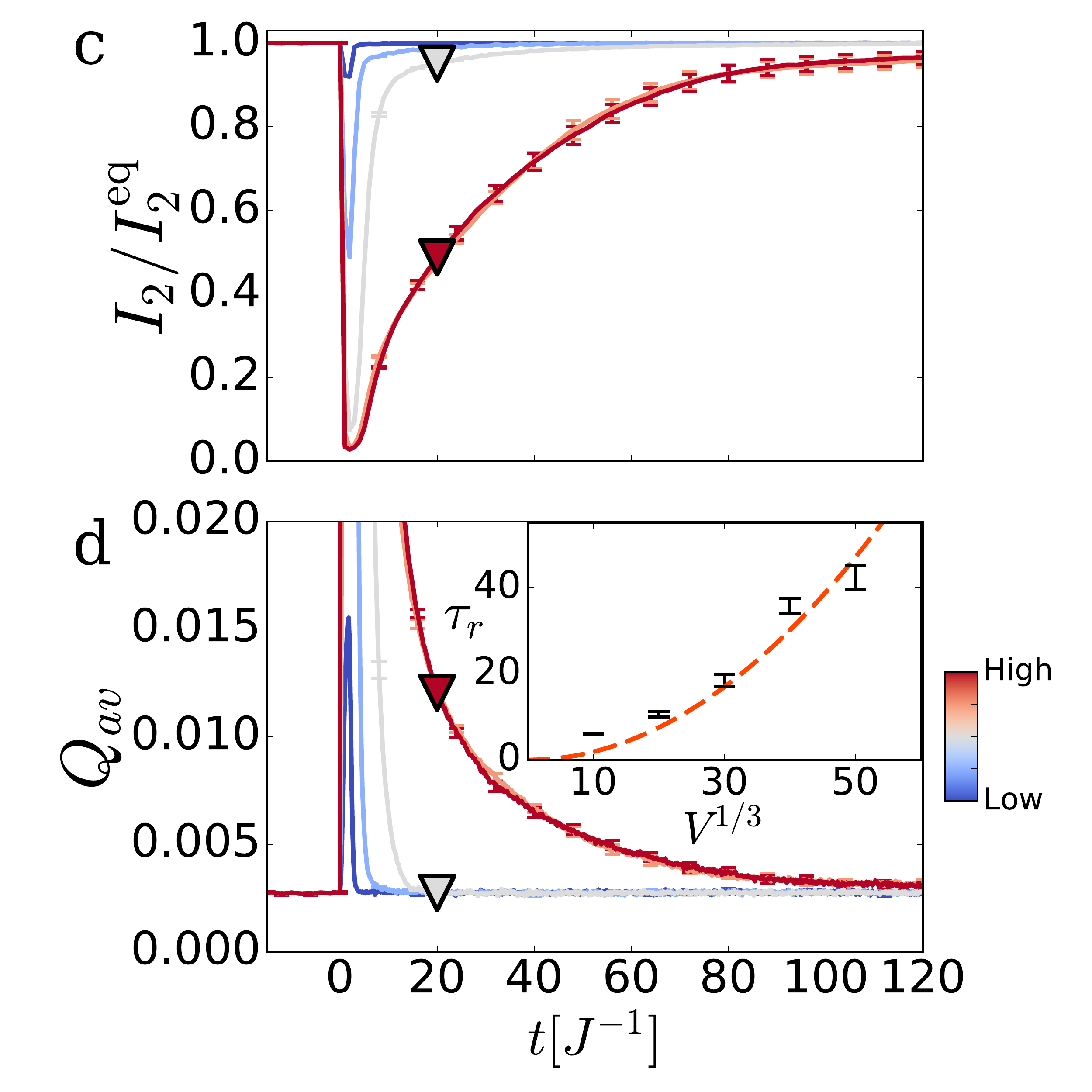}};
    \end{tikzpicture}
    \caption{(a,b) Snapshots of vorticity for
two individual quenches at time
$t=20 J^{-1}$:
(a)~weak quench, (b)~strong quench, marked in panels (c,d)
by gray and red triangles, respectively  [also shown in 
Fig.~\ref{quench:fig:experiment_CDW}(b)]. Vorticity $\mathbf{w}_j$ 
is depicted by (colored) arrows whose length is proportional to
$|\mathbf{w}_j|$
(see the text). For the visualization purposes, we apply a smoothing filter to $\mathbf{w}_j$: at site $j$ we plot $\mathbf{w}_j$ averaged over $9$ sites comprising a cube with side $3$ encompassing site $j$. The arrow color represents 
the position along one of the lattice axes, 
as shown by the color bar. 
In~(a) these arrows are visible as mere dots due to smallness of
$|\mathbf{w}_j|$.
In~(b) longer and hence brighter  arrows merge into distinctly visible worm-like bundles revealing
vortex cores. Vortex lines penetrating the entire system are clearly seen for
the stronger quench. (Further snapshots of vorticity for the later times of the stronger quench can be found in the Supplement \cite{supplemental}.)
\label{quench:fig:snapshots_of_defects_18}
(c,d) Correlation between vorticity and the slowing down of the order recovery for the simulated 3D DGPE lattice.
(c)~the 2D-integrated intensity $I_2(t)$, same as Fig.~\ref{quench:fig:experiment_CDW}(b) but over a longer time interval;
(d)~the average vorticity
$Q_{av}(t)$.  The color-coding and the statistical ensemble behind the sampling are used the same as in Fig.~\ref{quench:fig:experiment_CDW}(b).
\label{quench:fig:defects}
(d, inset) Dependence of the order parameter recovery time $\tau_r$ on the linear lattice size $V^{1/3}$.
\label{quench:fig:scaling_recovery_time}
}
\end{figure}

{ \it Simulations of non-equilibrium quenches. ---}
According to Ref.~\cite{dolgirev2020amplitude},
the energy deposited by the laser pulse into LaTe$_3$ is first absorbed by
the electrons, then transferred partially to the ``hot phonons", drastically
increasing their temperature and thereby melting the CDW order. After that,
the energy of the hot phonons is transferred to the rest of the
phonons, which have significantly larger specific heat. This
brings the overall temperature of the system to nearly the same temperature
as the one before the pulse. 
The phase degrees of freedom we aim at describing with the help of the
DGPE are supposed to belong to hot phonons. We model the above process by first
quickly pumping the energy into the DGPE lattice (heating of hot phonons by
electrons) and then quickly removing the energy from the lattice (cooling
of hot phonons by the energy transfer to the main phonon bath). 
Technically the above scheme is implemented by temporarily adding the
energy-nonconserving terms to the right-hand-side of
Eq.~(\ref{Intro:eq:equations-motion-DGPE})
as explained in
Refs.~\cite{tarkhov2020ergodization, tarkhov2021transient, supplemental}.

We simulated a set of quenches starting from the equilibrium state at $T \approx 0.4T_c$ and then using varying amounts of energy deposited to and then removed from the system, and averaged over $40$ random initial conditions, thermalized before the quench for $t = 600 J^{-1}$~\cite{supplemental}. The time dependencies of energy during the quenches is shown in the inset of
Fig.~\ref{quench:fig:experiment_CDW}(b).
These quenches imitated the
experimental laser pulses of different fluences used to produce the
experimental plots in
Fig.~\ref{quench:fig:experiment_CDW}(a). 
 
To compare our simulation with the diffraction
experiment~\cite{zong2019evidence},
we define the two-dimensional integrated spectral weight
$I_2 (t)$
by the following equality
\begin{eqnarray}
\label{eq::I2_def}
I_2 = \sum_{k_x, k_y}
        |\psi (k_x, k_y, 0)|^2,
\end{eqnarray}
where 
$\psi ({\bf k}) = V^{-1} \sum_j \psi_j e^{i {\bf k} \cdot {\bf r}_j}$
is the Fourier transform of 
$\psi_j$.
Summation
in
Eq.~(\ref{eq::I2_def})
runs over quantized momentum projections
$k_{x,y}$
lying in the Brillouin zone, while
$k_z = 0$.
Definition~(\ref{eq::I2_def})
mimics the 2D-integrated diffraction intensity measured experimentally in
Ref.~\cite{zong2019evidence},
see
Fig.~\ref{quench:fig:experiment_CDW}\,(a).
The square of the order parameter amplitude
$|\Psi|^2$
is the largest contributing term in the 
sum~(\ref{eq::I2_def}).
However, 
$I_2 > |\Psi|^2$
since shorter-range order fluctuations enhance 
$I_2$
relative to 
$|\Psi|^2$.
Thus, one can think of 
$I_2$
as a quantity tracking both long-range and short-range correlations in the DGPE system. We have further verified~\cite{supplemental} that $I_2(t)$ is proportional to the correlation length of the CDW order, $l_c(t)$, as anticipated in Refs.~\cite{zong2019evidence,dolgirev2020amplitude}.

The main panel of
Fig.~\ref{quench:fig:experiment_CDW}(b),
presents 
$I_2 (t)$
before, during and after the quench. The comparison between the simulation and the
experiment reveals a remarkable agreement: At small intensities of the quench, the spectral
intensity quickly recovers its pre-quench equilibrium value on the time scale comparable to the
duration of the quench itself. As the quench intensity grows, the character of the dynamics
undergoes a qualitative change: 
$I_2$
drops to zero and stays low for finite time interval. This happens when, during the quench,
the energy crosses the critical value
$\varepsilon_c$ 
associated with the phase transition and indicated by the dashed line in the inset of
Fig.~\ref{quench:fig:experiment_CDW}(b).
Most importantly, the post-quench recovery is extremely slow: the corresponding curves
visually split away from the curves representing weaker quenches.  

{ \it Topological defects. ---}
Here we investigate the possible connection between the slowdown of the order recovery and
the dynamics of topological defects. For 3D DGPE systems with periodic boundary conditions,
the topological defects form closed-loop
vortices~\cite{rojas2001kinetics}.
In this respect, the 3D DGPE model is  similar to the three-dimensional
$O(2)$,
$U(1)$
and XY-models, where the percolation of vortex loops is responsible for a
phase transition~\cite{hulsebos1994behavior, kajantie20002, kohring1986role}.

To monitor the local topological charge (vorticity), we define the dual lattice by translating the original lattice by vector
$(1/2,1/2,1/2)$.
Each of the sites of the dual lattice is thus surrounded by
$6$ square
plaquettes of the original lattice.

As an indicator of vorticity we use the finite-difference counterpart $\mathbf{w}_j$ of the continuous curl of the
current
\mbox{$(\nabla \times \mathbf{v}_s)_j$},
where
$\mathbf{v}_s = -i (\psi^* \nabla \psi - \psi \nabla \psi^*)$ (see ~\cite{supplemental}).
In
Fig.~\ref{quench:fig:snapshots_of_defects_18}(a,b),
we provide a space-resolved snapshots of field $\mathbf{w}_j$ 
generated in the course of the quenches shown in
Fig.~\ref{quench:fig:experiment_CDW}(b). The snapshots correspond to the two states marked in
Fig.~\ref{quench:fig:experiment_CDW}(b)
by red and gray triangles.
In principle, each site of the dual lattice has a non-zero vorticity $\mathbf{w}_j$, which is the consequence of the discreteness of the lattice field. However, in the absence of a real vortex, such a vorticity is very small and does not form any line pattern, while the presence of a vortex is indicated by the large values of $\mathbf{w}_j$ showing the vortex core as a colored ``tube'' extended through the lattice. 

Fig.~\ref{quench:fig:snapshots_of_defects_18}(a)
illustrates a weak quench, for which the order (quantified by
$I_2$)
recovers quickly. 
One can see that  the system does not exhibit any colored tube indicative of a vortex.  For more intensive quenches, on the contrary,
one can observe vortex tubes percolating through large volumes of the system and forming
clearly identifiable large vortex loops,
see Fig.~\ref{quench:fig:snapshots_of_defects_18}(b) (see also Fig.~S.7(a-c)
and the video in the supplemental material~\cite{supplemental}).
As one can see, for example, in the video, these loops initially form an entangled network.
The crossings of the loop lines then lead to smaller loops separating from this network and
collapsing, while one large loop spreading over most of the lattice is eventually being formed.
The time required for that last loop to shrink and collapse is what determines the time scale of the
order recovery in
Fig.~\ref{quench:fig:experiment_CDW}(b).   


We also introduce an auxiliary binary vorticity variable $Q_j$ for each dual lattice site~\cite{supplemental}: it is defined such that 
$Q_j =1$, if at least one of the 6 adjacent original-lattice plaquettes has non-zero vorticity; otherwise $Q_j =0$.
Finally, we define the average vorticity for the entire system as  $Q_{av}  = \frac{1}{V} \sum_j^V Q_j$.

Further evidence that the relaxation of vorticity
$Q_{av}(t)$ is directly correlated with the order recovery after a quench is presented in Figs.~\ref{quench:fig:defects}(c,d): there
the strong quenches exhibiting the slowdown of the $I_2$
recovery are accompanied by pronounced slowly relaxing tails of
$Q_{av}(t)$.


In
Fig.~\ref{quench:fig:scaling_recovery_time}(c, inset),
we show that the order parameter recovery $\tau_r$, defined as the time to
reach $75\%$ of the initial equilibrium value
of $|\Psi|^2$
for the strongest quench, grows with the lattice volume (number of lattice sites) $V$. The
simulations are consistent with the scaling
$\tau_r \propto V^{2/3}$.
We have not attempted to derive this power law. Yet, the growth of
$\tau_r$ 
with the lattice size as such indicates that the recovery of the order is contingent on the
disappearance of the extensive vortex loops that penetrate the entire system and need to drift by the distances of the order of the lattice size
before they annihilate.

{ \it Conclusions ---} In conclusion, our direct simulations of the DGPE lattice with a CDW-like
form of the order parameter are consistent with the conjecture of
Ref.~\cite{zong2019evidence}
that the observed slowdown of the recovery of the CDW order after
laser-induced melting is due to the emergence and then slow disappearance of topological defects
in the order parameter texture. We were able to numerically monitor the vorticity in the system
and thereby establish the correlation between the relaxation of vorticity and the recovery of
the order. In a broader context, our investigation illustrates the viability of using DGPE for
simulating the CDW dynamics even though DGPE as such was historically introduced to describe
superfluid systems. 

{ \it Code availability ---} The source code is published in a GitHub repository~\cite{githubrepo}.

{\it Acknowledgments. ---} This work was supported by a grant of the Russian Science Foundation (Project No. 17-12-01587). The work at MIT was mainly funded by the U.S. Department of Energy, BES DMSE (data taking and analysis) and the Gordon and Betty Moore Foundation’s EPiQS Initiative grant GBMF9459 (instrumentation).



\bibliographystyle{apsrev4-1}
\bibliography{refs3}

\newpage

\renewcommand{\theequation}{S.\arabic{equation}}
\setcounter{equation}{0}

\renewcommand{\thefigure}{S.\arabic{figure}}
\setcounter{figure}{0}

\section{Supplemental Material} 

\subsection{S.I. Details of the  experiment of
Ref.\cite{zong2019evidence}}

Here, for the convenience of the readers, we provide additional details of the experiment of
Ref.\cite{zong2019evidence}. 

\begin{figure}[h]
    \centering
    \includegraphics[width=0.75\columnwidth]{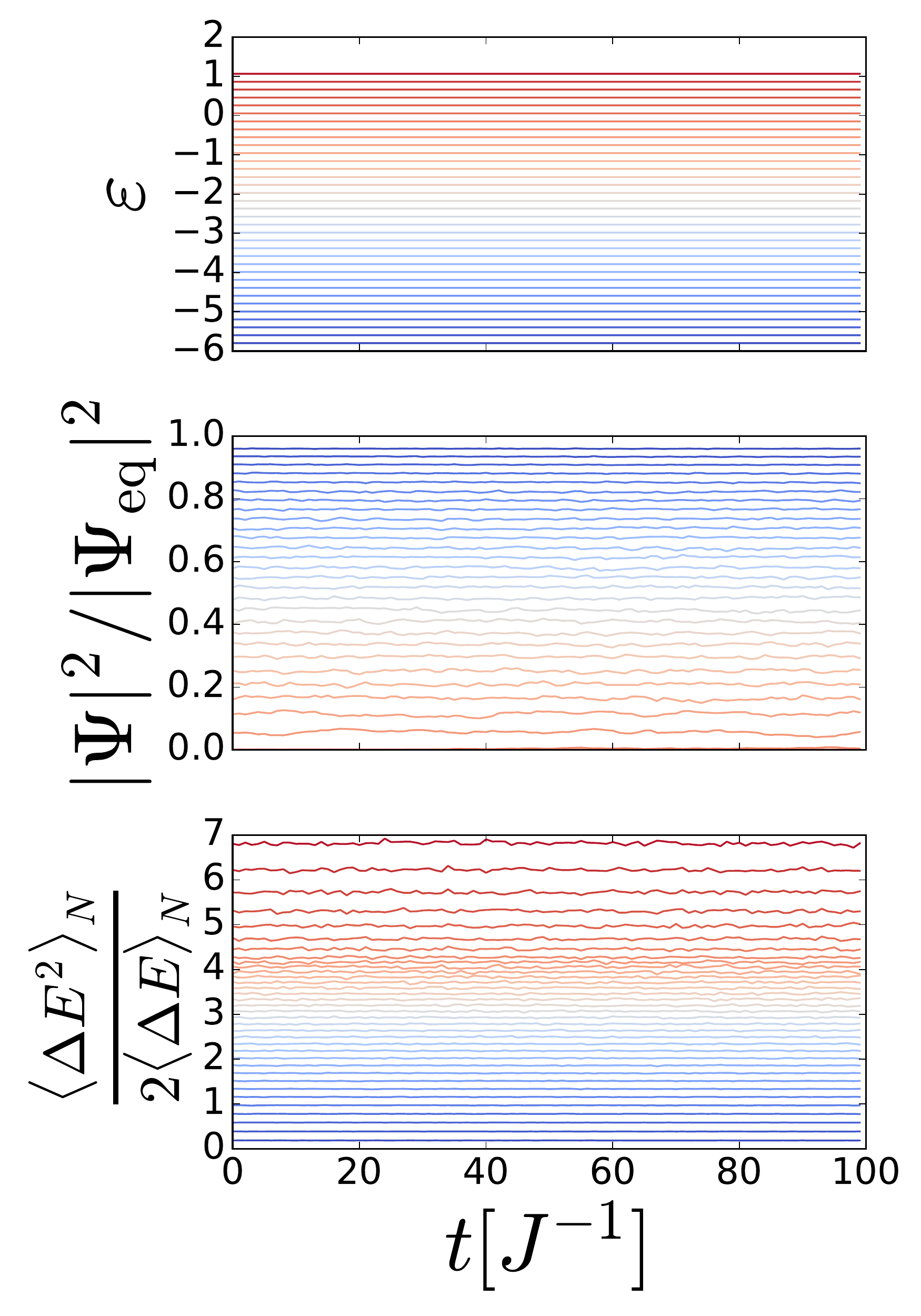}
    \caption{Thermalization dynamics behind Fig.~\ref{quench:fig:thermodynamics}, $V=50^3$: energy density, order parameter and the right-hand-side of  Eq.~(\ref{Intro:eq:T_E_fixed_N})}\label{quench:fig:equilibrium_thermalization}
\end{figure}

The material studied, LaTe$_3$,
has a layered crystal structure consisting of alternating Te bilayers and LaTe layers. The critical temperature of the CDW transition in LaTe$_3$
is about
$700$~K. The CDW is localized primarily in the Te bilayers. 

In the experiment of Ref.\cite{zong2019evidence}, femtosecond laser pulses were applied to LaTe$_3$ and then the CDW response was probed by (i) optical reflectivity measurements, (ii) time-resolved angle-resolved photoemission spectroscopy and (iii) the ultrafast electron diffraction. 
Fig.~\ref{quench:fig:experiment_CDW}(a) of the main text shows the results for the latter technique, namely, the 2D-integrated intensity, $I_2$, of the CDW electron diffraction peak is
plotted as a function of delay time for various laser pulse fluences. The peak intensity is integrated for the two in-plane wave-vector
projections. The intensity is interpreted as being proportional to the
square of the CDW order-parameter multiplied by the CDW correlation length
for the out-of-plane direction (the one perpendicular to the Te layers). 


By examining the time evolution of the width of the CDW peak, the
authors Ref.\cite{zong2019evidence}  provided indirect evidence that the
slow recovery of the intensity of the CDW diffraction peak seen in
Fig.~\ref{quench:fig:experiment_CDW}(a)  is caused by the emergence of
topological defects of the CDW order, followed by their subsequent slow
annihilation. Even though such a scenario is rather natural, no direct evidence in real space in its favor was presented.

\subsection{S.II. Microcanonical thermodynamics of the DGPE lattice: Numerical aspects}

\subsection{Equilibrium properties}

For the parameters used in the main text, the solutions of DGPE with fixed $E$ and $N$ were observed to exhibit the behavior consistent with the ergodic hypothesis. Namely, the right-hand-side of Eq.~(\ref{Intro:eq:T_E_fixed_N}) computed for a given phase space trajectory and used below  to define the temperature was found to converge to a value independent of the initial conditions.

To describe the statistical properties of the DGPE lattice, we use the standard definition of the microcanonical temperature:
\begin{equation}
\label{Intro:eq:microcan_T_def}
\frac{1}{T} \equiv \frac{\partial S}{\partial E},
\end{equation}
where
$S \equiv \text{log} \, W(E)$
is the microcanonical entropy, with  $ W(E)$ being
the  $(V-2)$-dimensional volume of the energy shell at constant $N$. We set the Boltzmann constant 
$k_{\rm B} = 1$.

\begin{figure}[h]
    \centering
    \includegraphics[width=0.75\columnwidth]{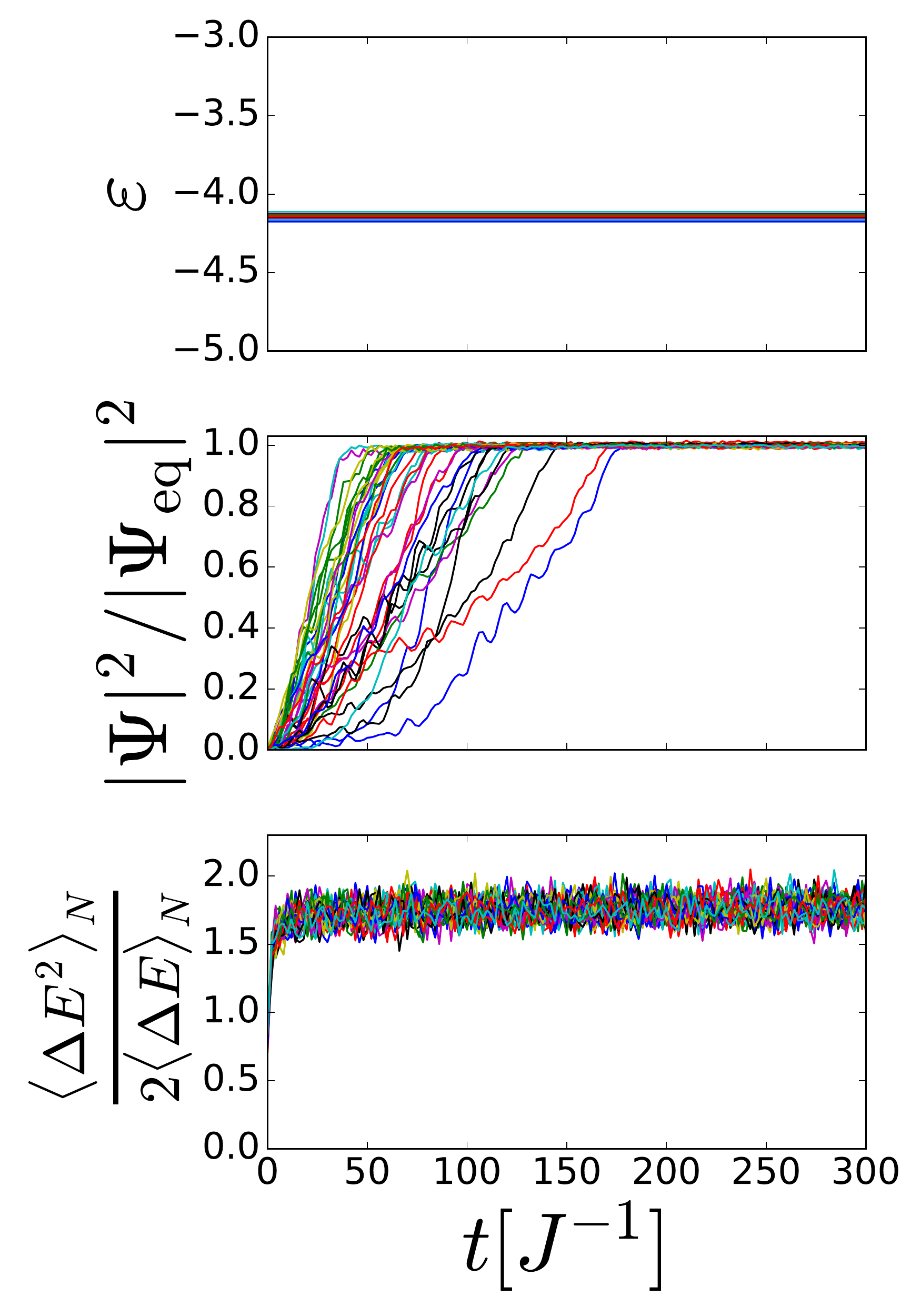}
    \caption{Thermalization of pre-quench states for $40$ on-shell initial conditions generated  by the procedure described in Section~S.II with $V=50^3$.}\label{quench:fig:thermalization}
\end{figure}

The direct determination of the volume $ W(E)$ in the many-dimensional phase space is a computationally prohibitive task. It is, however, possible to use Monte-Carlo sampling to determine directly $\frac{1}{W(E)} \frac{d W(E)}{dE}$, which is equal to the right-hand-side of Eq.(\ref{Intro:eq:microcan_T_def}).
We do this by adapting the numerical recipe developed for classical spin lattices~\cite{de2015chaotic} and consistent with the analytical
approach of Refs.~\cite{rugh1997dynamical, rugh1998geometric}. The idea is based on the observation that the rate of growth of a many-dimensional energy shell is determined by the average local curvature of that shell. The latter, in turn, can be locally sampled in the vicinity of each point by introducing small isotropic random perturbations to the phase point coordinate and then computing the variance of energy fluctuations
$\langle\Delta E^2\rangle$
and the mean fluctuation
$\langle\Delta E\rangle$.
We note that
$\langle\Delta E\rangle \neq 0$
due to the fact that, in general, there are exponentially more states above
the energy shell than below it. 
We use the above approach  with 
the additional constraint that the small perturbations  should leave $N$ constant. The corresponding average is denoted as $\langle \ldots \rangle_N$.
With such a notation, the numerically computed expression for temperature becomes 
\begin{equation}
\frac{1}{T} = \frac{2\langle\Delta E\rangle_N}{\langle\Delta E^2\rangle_N}.
\label{Intro:eq:T_E_fixed_N}
\end{equation}
In the limit $V \to \infty$, computing the averages entering Eq.(\ref{Intro:eq:T_E_fixed_N})  in the vicinity of one randomly chosen sample point on an energy shell should be sufficient to determine the temperature. However, since our numerically simulated lattice is finite, we supplement the above procedure by the additional averaging of the right-hand-side of Eq.(\ref{Intro:eq:T_E_fixed_N}) over points along a randomly chosen phase space trajectory and then further over phase space trajectories with randomly chosen initial conditions. 

\begin{figure}[h]
    \centering
    \includegraphics[width=0.6\columnwidth]{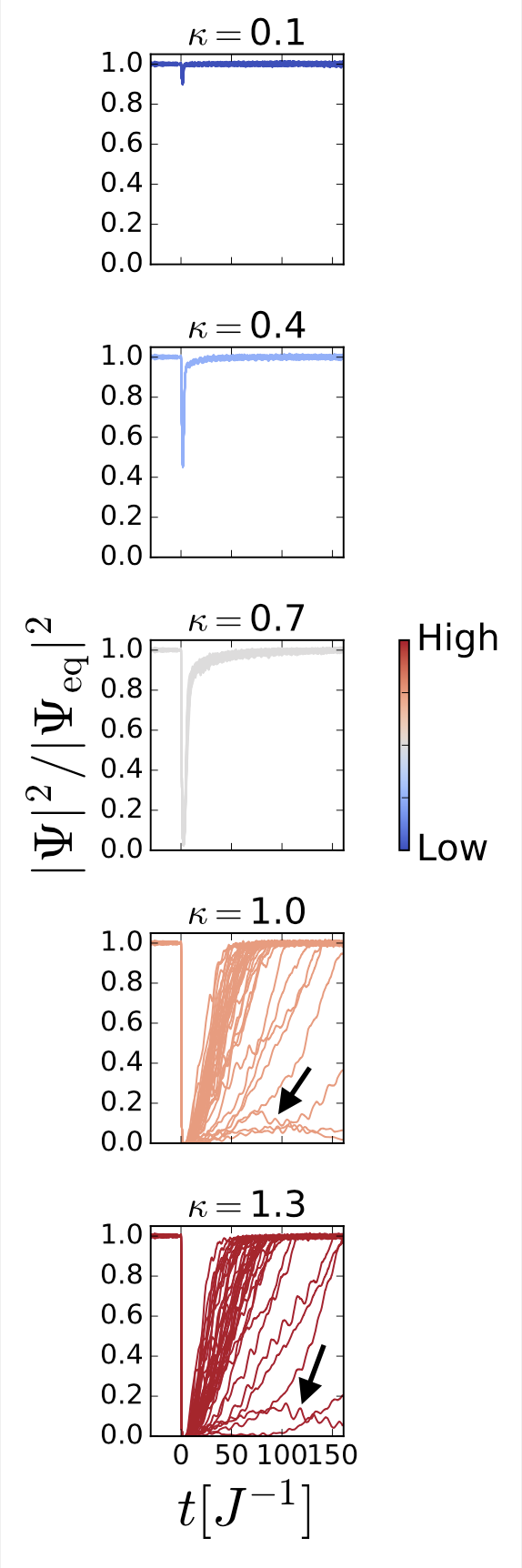}
    \caption{Quench realizations behind the averages plotted in Figs.~\ref{quench:fig:experiment_CDW}(b) and \ref{quench:fig:defects}(a) -- $40$ initial conditions, $V = 50^3$.  For stronger quenches $\kappa = 1.0$ and $\kappa = 1.3$, atypical slowly relaxing solutions appear -- they indicate the presence of large topological defects in the lattice. We do not always observe the annihilation of defects, in some cases the transient resurgence of vorticity occurs (black arrows indicate such quench realizations).
    \label{quench:fig:individual_quenches}
    }
\end{figure}

Technically, the calculation of equilibrium properties plotted in Fig.~\ref{quench:fig:thermodynamics} consists of the following steps: 

\begin{enumerate}
    \item[(i)] choosing the initial configurations $\{\psi_j \}$ as a random Hilbert-space vector normalised by the condition $\sum_j |\psi_j|^2 \equiv N = V$.    This state has randomly chosen phases and amplitudes, and corresponds to $\varepsilon \sim 0$ and $T \sim T_c$; 
    \item[(ii)] choosing an energy value for which the temperature is to be calculated; 
    \item[(iii)] activating the non-conservative term in the DGPE Eq.~(\ref{eq::dgpe_quench}) with function $K \equiv \const$ for a time interval sufficient to reach a desired energy value (see Section S.III); 
    \item[(iv)] deactivating the non-conservative term, then calculating the conservative dynamics of the DGPE for a sufficiently long time $t = 100 J^{-1}$ (see Fig.~\ref{quench:fig:equilibrium_thermalization}); 
    \item[(v)] choosing the next slightly smaller target energy value and then, starting from the thermalized state at the final time point from the preceding step, repeating steps (ii) to (v).
\end{enumerate}
The overall procedure guarantees that the DGPE lattice has enough time to thermalize at each energy shell. 

\subsection{Preparation of the initial state preceding the quenches.}

In Fig.~\ref{quench:fig:thermalization}, we illustrate the preparation of the initial thermalized state for the quenches presented in Fig.~\ref{quench:fig:experiment_CDW}(b) of the main text of the paper. We start from the high-temperature state  of step (i) above and then quickly drive the system to $\varepsilon = -4.1$ corresponding to $T \approx 0.4T_c$. Such a driving initially produces a strongly nonequilibrium state (in contrast to the slow preparation procedure used in the previous subsection). Then, we track the thermalization dynamics of $\varepsilon$, $\frac{|\Psi|^2}{|\Psi_{\rm{eq}}|^2}$ and $\frac{2\langle\Delta E\rangle_N}{\langle\Delta E^2\rangle_N}$ from Eq.~(\ref{Intro:eq:T_E_fixed_N}). We observe that the energy is conserved, the order parameter reaches its equilibrium value by $t \sim 150 J^{-1}$, while the parameter $\frac{2\langle\Delta E\rangle_N}{\langle\Delta E^2\rangle_N}$ starts exhibiting the typical behavior faster, namely, after times of the order of $t \sim 10 J^{-1}$. By  time $t = 200 J^{-1}$, all the parameters reach their equilibrium regimes independently of the initial conditions. For the quenches, presented in Fig.~\ref{quench:fig:experiment_CDW}(b) of the main text, the conservative dynamics were run for $t = 600 J^{-1}$, which is significantly longer than required for equilibration and thermalization. Hence, before the start of the quenches in Fig.~\ref{quench:fig:experiment_CDW}(b) the system was well-thermalized, and the initial conditions for the quenches are typical for the energy shell.


\subsection{S.III. Numerical implementation of the DGPE lattice quench\label{quench_suppl}}

The quench is performed on the DGPE lattice prepared in equilibrium at
$\varepsilon = -4.1$,
$T/T_c \approx 0.4$.
During the quench, the system's energy experiences quick rise and fall
within a limited time interval.
To allow for the time variation of $E$, we
modify~\cite{tarkhov2020ergodization, tarkhov2021transient,kamaletdinov2021vortex_annealing}
DGPE by adding a non-energy-conserving term (the last one below):
\begin{eqnarray} 
\label{eq::dgpe_quench}
\nonumber i\frac{d\psi_{j}}{dt}
=
-\sum_{k\in \text{NN}(j)}
	\psi_{k} + \frac{Un}{J} \left|\psi_{j}\right|^{2}\psi_{j} - \\
	 - K(t) \psi_{j} D_j,
\end{eqnarray}
where
\begin{eqnarray}
D_j^{\vphantom{*}} = D_j^* = i \sum_{k\in \text{NN}(j)}
\left( \psi_{k}^* \psi_{j} - \psi_{k} \psi_{j}^* \right),
\end{eqnarray} 
 and the function 
$K(t, \kappa)$ is to be specified below.
Note that
Eq.~(\ref{eq::dgpe_quench}),
remains ``gauge invariant'' and thus conserves the norm $N$.

To verify the energy-nonconserving character of the added term one can use 
Eq.~(\ref{eq::dgpe_quench}) to obtain
\begin{equation}
    \dot{E}  = - K(t) \sum_j D_j^2.
\end{equation}
The sign of
$K(t)$ 
controls the increase or decrease of $E$: function $K$ must be
negative to heat the system, and positive to cool it. We
additionally require that the energies of the system before and after the
quench are identical. 

All our simulated quenches use the function
$K(t)$ having the following form
\begin{eqnarray} 
\label{quench:eq:kappa_profile}
K(t)
&=& - \theta (t) \left[
	\ \kappa \ \frac{\gamma_2 e^{-\gamma_2 t} - \gamma_1 e^{-\gamma_1 t}}
		{\gamma_2 - \gamma_1}
	\right.
\\
\nonumber 
	&-&	
	\left.
	\theta (t-t^*) \gamma_3 \left(1 - e^{-\gamma_2 (t-t^*)}\right)
	\frac{E(t)-E_0}{E^*-E_0}
\right],
\end{eqnarray} 
where
$\theta (t)$
is the Heaviside step-function,
$\kappa$
is the parameter controlling the
strength of the quench (in our simulations, $\kappa$ spans  the range from 0.1 to 20), $\gamma_1=1$,
$\gamma_2=0.3$ and $\gamma_3=0.01$ are numerical parameters,
\begin{eqnarray} 
t^*=\frac{1}{\gamma_1-\gamma_2}
	\log \left(\frac{\gamma_1}{\gamma_2} \right)
\end{eqnarray} 
is the time at which
$K(t)$
changes sign,
$E_0$
is the pre-quench energy of the system, 
$E(t)$
is the energy at time
$t$,
and
$E^*=E(t^*)$.

\begin{figure}[h]
    \centering
    \includegraphics[width=0.75\columnwidth]{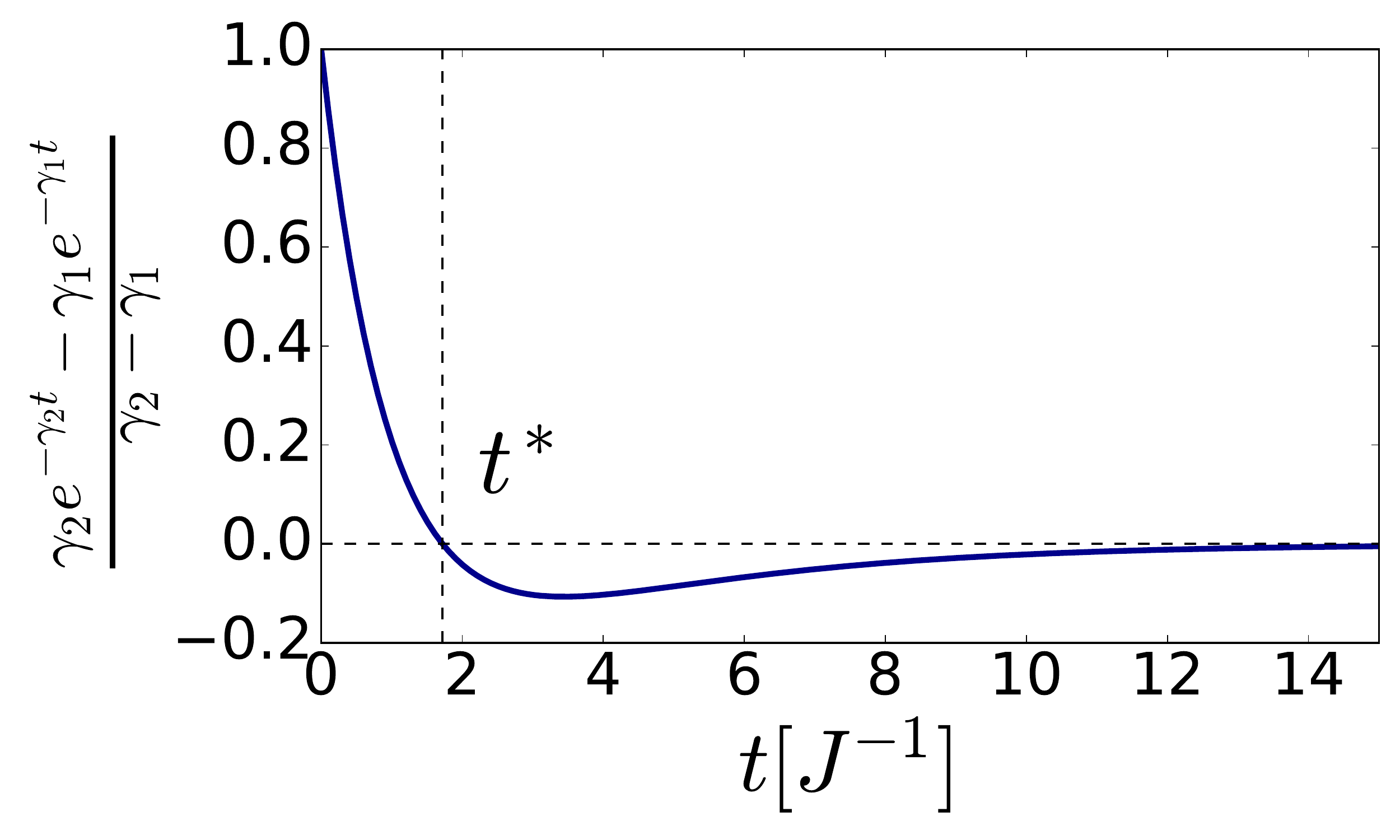}
    \caption{The principal factor in the first term in  Eq.~(\ref{quench:eq:kappa_profile}) as a function of time.}\label{quench:kappa_first_term}
\end{figure}

The quench is principally determined by the first term in the square brackets in Eq.(\ref{quench:eq:kappa_profile}) -- the time-dependence of this term is shown in Fig.~\ref{quench:kappa_first_term}. 
The second term in
the square brackets
is added to guarantee that the energy after the quench is equal to the initial energy $E_0$.

Before the quench, the system is prepared in an equilibrium low-energy
state as described in Section S.II. At the first stage of the quench the system is subjected to the fast
energy increase (heating). The duration that stage is $t^*$.
The heating is then followed by the cooling stage of duration regulated by parameters  
$\gamma_2$ and $\gamma$. The resulting time evolution of the energy density
$\varepsilon (t)$ for various quench parameters $\kappa$
is plotted in the inset of Fig.~\ref{quench:fig:experiment_CDW}(b) of the main text.

\begin{figure}[h]
    \centering
    \includegraphics[width=0.9\columnwidth]{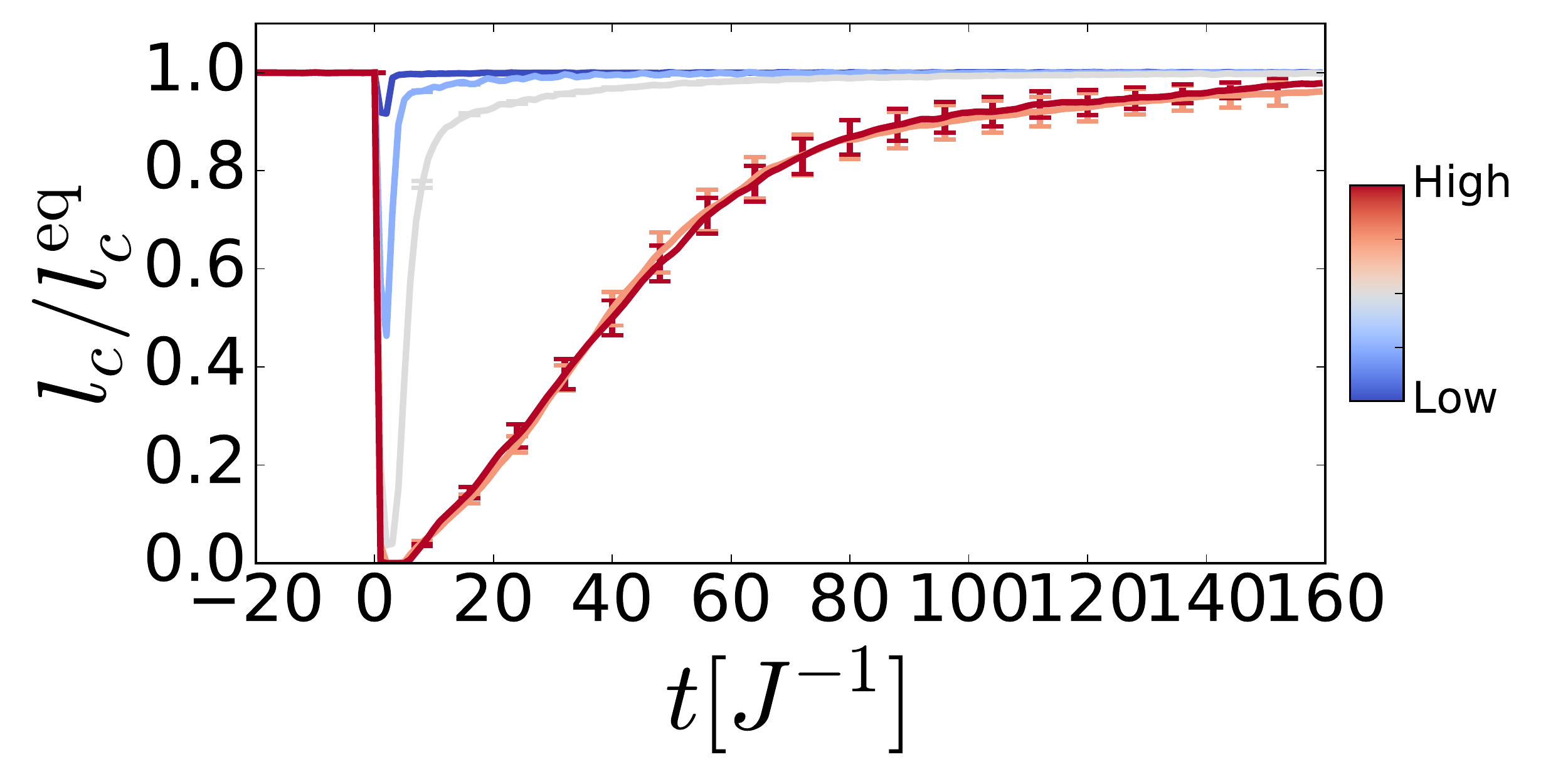}
    \caption{Time dependence of the correlation length for the quench illustrated in Fig.~\ref{quench:fig:experiment_CDW}(b) and  Fig.~\ref{quench:fig:defects}(c).}\label{quench:correlation_length}
\end{figure}

\begin{figure}[h!]
    \begin{tikzpicture}
    \node[inner sep=0pt] (duck) at (0,0)
    {\includegraphics[width=\columnwidth]{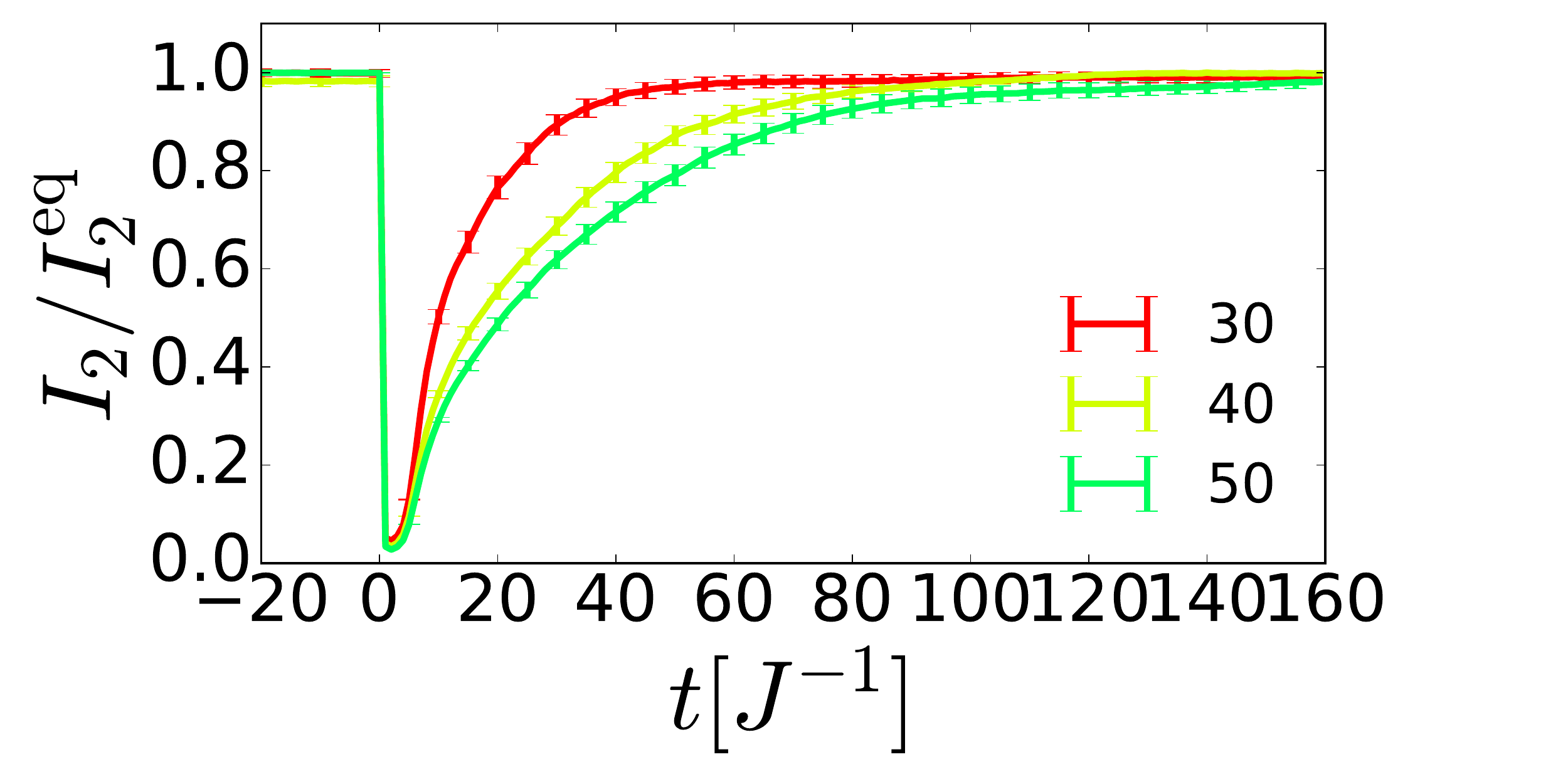}};
    \node[align=center,fill=white] at (-3.8, 1.8) {\textbf{a}};
    \end{tikzpicture}
    ~
    \begin{tikzpicture}
    \node[inner sep=0pt] (duck) at (0,0)
    {\includegraphics[width=\columnwidth]{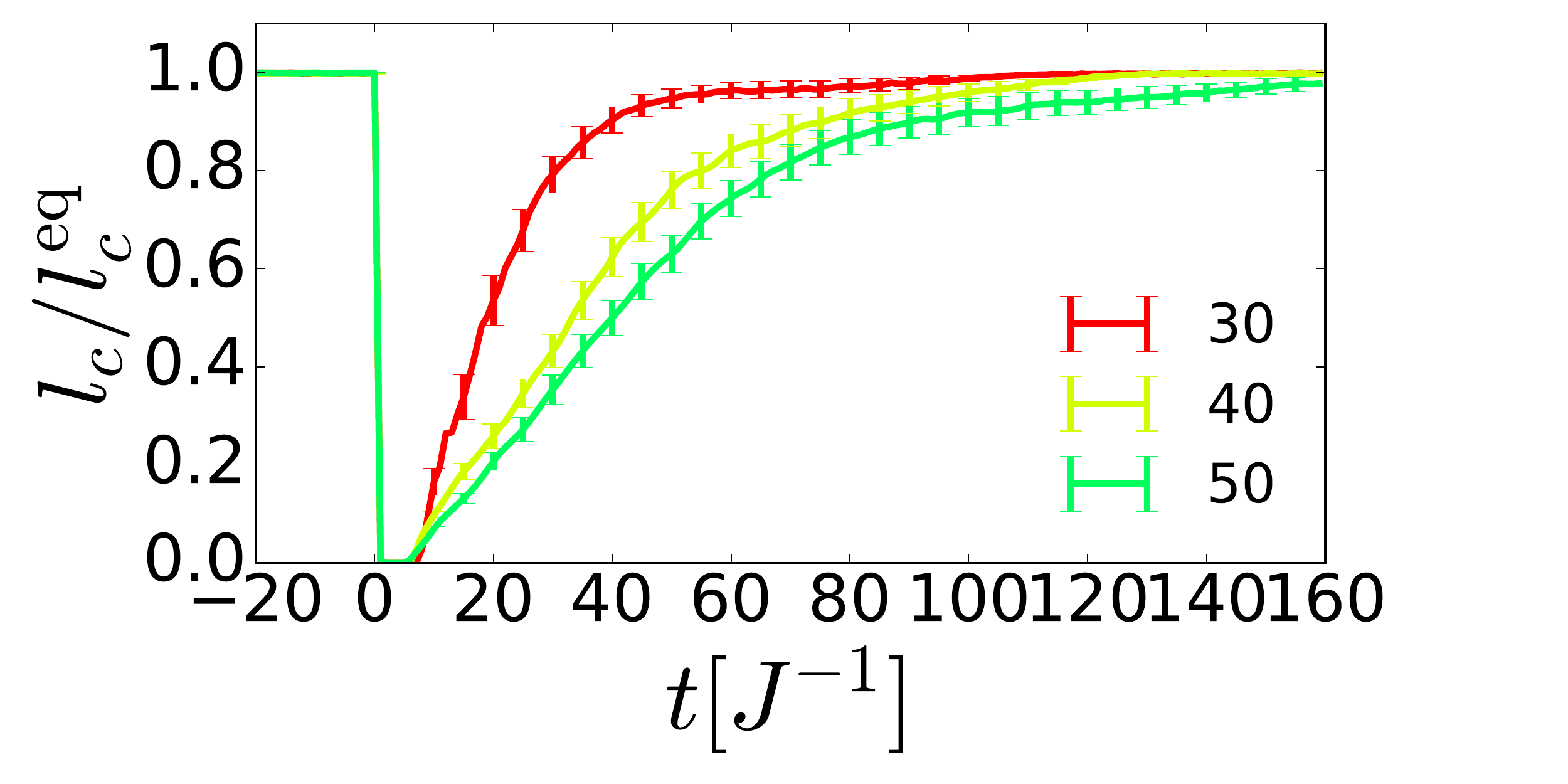}};
    \node[align=center,fill=white] at (-3.8, 2.0) {\textbf{b}};
    \end{tikzpicture}
\caption{System size scaling for the time evolution of (a) the theoretically computed post-quench recovery, $I_2$, and (b) the correlation length, $l_c$. The color of the curves represents
different system sizes.
\label{quench:fig:scaling_Lcorr_I2}
}
\end{figure}

For further details one can refer to the source code published in a GitHub
repository~\cite{githubrepo}.

\subsection{S.IV. Estimate of parameter $U$ for the DGPE lattice}


In order to estimate the nonlinearity parameter $U$, we match the 
velocity of the phasons propagating within the real CDW system and the
velocity of acoustic excitations calculated theoretically for the DGPE model. To
perform the desired calculation, we treat phasons as weak perturbations
distorting spatially homogeneous zero-temperature solution
$\Psi_0 (t)$.
Within such an approach, the original DGPE variables
$\psi_j$
should be expressed as the sum
$\psi_j(t) = \Psi_0 (t)+ \xi_j(t)$,
where variables
$\xi_j(t)$
are small. The DGPE now reads
\begin{eqnarray} 
\label{quench:eq:DGPE_real_space_perturbation_expansion}
i\dot{\Psi}_0 + i\dot{\xi}_{j}
=
-2dJ \Psi_0 -J\sum_{m\in \text{NN}(j)}\xi_m +
\\
\nonumber + U \left|\Psi_0 + \xi_j\right|^2 \left(\Psi_0 + \xi_j\right).
\end{eqnarray} 
Neglecting all terms containing
$\xi$,
we derive that the unperturbed zero-temperature solution 
$\Psi_0 (t)$
must satisfy
\begin{equation}
i\dot{\Psi}_0 = -2dJ \Psi_0 + U \left|\Psi_0\right|^2 \Psi_0.
\end{equation}
The solution to the latter equation can be trivially found
\begin{equation}
\label{eq::suppl::Psi0}
\Psi_0 (t) = \sqrt{n_0} \exp{(-i t \mu)},
\end{equation}
where $n_0=|\Psi_0|^2$
and the ``chemical potential''
$\mu=-6 J + U n_0$.

When the deviations from the homogeneous zero-temperature
solution~(\ref{eq::suppl::Psi0})
are small
($|\xi| \ll \sqrt{n_0}$),
it is permissible to keep only the leading-order
terms in
Eq.~(\ref{quench:eq:DGPE_real_space_perturbation_expansion}).
This allows us to derive the following linear differential equation
\begin{equation}
\label{quench:eq:DGPE_real_pert_no_time_dependence_linear_approx}
i\dot{\zeta}_{j}
=
-J\sum_{m\in \text{NN}(j)}\zeta_m -\mu \zeta_j
+
U\left( 2 n_0\zeta_j + n_0 \zeta_j^*\right).
\end{equation}
The new dynamical variables
$\zeta_j$
introduced in this equation are connected to
$\xi_j$
by the relation
$\xi_j =\zeta_j e^{-i t \mu}$.

In Fourier space
Eq.~(\ref{quench:eq:DGPE_real_pert_no_time_dependence_linear_approx})
becomes
\begin{equation}
\label{quench:eq:DGPE_fourier_space_time_dep}
i\dot{\zeta}_{\mathbf{k}}
=
(\varepsilon_\mathbf{k} -\mu + 2 U n_0) \zeta_\mathbf{k}
+
U n_0 \zeta_\mathbf{-k}^*,
\end{equation}
where 
$\varepsilon_\mathbf{k}$
is defined as
\begin{equation}
\varepsilon_\mathbf{k}
=
-2J\left[ \cos{(k_x l)} + \cos{(k_y l)} + \cos{(k_z l)} \right].
\end{equation}
Since
Eq.~(\ref{quench:eq:DGPE_fourier_space_time_dep})
couples
$\zeta_{\bf k}$
and
$\zeta_{-\bf k}^*$,
we need to solve
Eq.~(\ref{quench:eq:DGPE_fourier_space_time_dep})
together with the equation for 
$\zeta_{-\bf k}^*$,
which reads
\begin{eqnarray} 
\label{quench:eq:DGPE_fourier_space_time_dep2}
-i\dot{\zeta}_{\mathbf{-k}}^*
=
(\varepsilon_\mathbf{k} -\mu + 2 U n_0) \zeta_\mathbf{-k}^*
+
U n_0 \zeta_\mathbf{k}.
\end{eqnarray} 
Solving the system of 
equations~(\ref{quench:eq:DGPE_fourier_space_time_dep})
and~(\ref{quench:eq:DGPE_fourier_space_time_dep2}),
we find the spectrum of excitations
$\omega_{\bf k}$
for the DGPE lattice. Frequencies
$\omega_{\bf k}$
satisfy
\begin{equation}
{\rm det}\,
\begin{pmatrix}
\label{quench:eq:dispersion_equation_for_diagonalization}
	\omega_\mathbf{k} -\varepsilon_\mathbf{k} + \mu - 2U n_0 & -U n_0\\
	- U n_0 & -\omega_\mathbf{k} -\varepsilon_\mathbf{k} + \mu - 2U n_0  
\end{pmatrix} = 0.
\end{equation}
The positive branch of the resulting dispersion relation is
\begin{equation}
\omega_\mathbf{k}
=
\sqrt{(6J+\varepsilon_\mathbf{k} + Un_0)^2 - U^2 n_0^2}.
\end{equation}
When
$l|{\bf k}| \ll 1$,
it can be further approximated as
\begin{equation}
\omega_\mathbf{k}
\approx
\sqrt{\left(J l^2 \mathbf{k}^2\right)^2 +  2 l^2 J U n_0 \mathbf{k}^2}
\approx
c_{\rm ex} | {\bf k} |,
\end{equation}
where 
$c_{\rm ex} = l \sqrt{2 J U n_0}$
is the speed of the acoustic excitations.

We map the DGPE onto the CDW dynamics by equating
$c_{\rm ex}$
with the speed of CDW phasons
$c_{\rm ph}$, thereby arriving to the relation:
\begin{equation}
\label{quench:eq:phasons_speed_DGPE}
c_{\rm ph} = l \sqrt{2 J U n_0}
\end{equation}
used in the main text.

We then use
Eq.~(\ref{quench:eq:phasons_speed_DGPE})
to estimate the DGPE nonlinearity term $U$ as follows:

\begin{eqnarray} 
\label{quench:eq:phason_velocity}
U = \frac{J}{2n} \left(\frac{c_{\rm ph}}{J l}\right)^2 = \frac{J}{2n} \left(\frac{c_{\rm ph}}{\Omega_{\rm D} a}\right)^2 = 
\\ 
\nonumber = (6 \pi^2)^{2/3} \frac{J}{2n} \left(\frac{c_{\rm ph}}{c_s}\right)^2,
\end{eqnarray} 
where
$c_s \sim (6 \pi^2)^{-1/3} \Omega_{\rm D} a$
is the speed of sound. 

As a further crude estimate, we now use
$c_{\rm ph} \sim c_s$,
thereby obtaining
$U \sim 10 \frac{J}{n}$.
(Ref.~\cite{manley2018supersonic}
showed that the CDW phason velocity can be
$2.8 - 4.3$
the sound velocity.)

 
\subsection{S.V. Definition of the vorticity characteristics}

The numerical calculation of the vorticity field
$\mathbf{w}_j$
for each dual lattice site is based on the fourth-order finite
differentiation scheme for each first-order derivative $\left(df(x)/dx\right)_j = (f_{j+2} -8 f_{j+1} + 8 f_{j-1} - f_{j-2})/12$. 

In order to define the vorticity variable $Q_j$, we first introduce the auxilliary variable $q_p$ for each plaquette surrounding the dual site $j$ as follows: (i) The DGPE lattice variables are parameterised as
$\psi_j= |\psi_j| \exp{(i \phi_j)}$. (ii) For a given plaquette consisting of vertices labelled as $1$,
$2$, $3$ and $4$, we define 
\begin{equation}
\begin{split}
q_{\text{p}} &
=
\frac{1}{2\pi} [\left.(\phi_{2} - \phi_{1} )\right|_{[-\pi, \pi)} +
	\left.(\phi_{3} - \phi_{2} )\right|_{[-\pi, \pi)} +   \\
 & + \left.(\phi_{4} - \phi_{3} )\right|_{[-\pi, \pi)} +\left.(\phi_{1} -
	\phi_{4} )\right|_{[-\pi, \pi)} ],
\end{split}
\label{quench:eq:def_vorticity}
\end{equation}
where the notation
$(...)|_{[-\pi, \pi)}$
indicates that the phase differences are ``backfolded" into the interval
$[-\pi, \pi)$.
The variable
$q_{\text{p}}$
can take only three possible values $-1$, $0$ and $1$. We then define the
variable
$Q_j$
such that
$Q_j = 0$
if all adjacent plaquettes
have~\mbox{$q_{\text{p}} = 0$},
and
$Q_j=1$
otherwise.

\subsection{S.VI. Relation between $I_2$ and the correlation length of the CDW order}

The $\delta$-function peak due to the CDW order is known to be broadened by the presence of topological defects, which make the peak to acquire a finite width. The intensity of the $\delta$-peak is then spread over a range of small wavevectors $q$, which in the real space then leads to a finite correlation length of the CDW order. The same is supposed to be true in our simulations around the ordering vector at $q=0$.

We have found empirically, that the low-$q$ intensity in the presence of vortices scales as \begin{equation}
    W(q) = A/q^2.
    \label{W}
\end{equation}
Such a scaling is naturally expected given that the low-$q$ fluctuations should be associated with the sound waves of a vortex liquid with dispersion $\omega \propto q$. The equilibrium thermal amplitude of each mode $a_q$  is supposed to obey the relation $a_q^2 \omega^2 = T$, from which Eq.(\ref{W}) follows given also that $W(q) \propto a_q^2 $. 

The singular character of the above shape precludes one from defining the correlation length from the half-width of the peak at half maximum. Instead, we define the correlation length as 
\begin{equation}
    l_c = \frac{\pi}{q_0},
    \label{lc}
\end{equation}
where $q_0$ is the upper cutoff value for $W(q)$ given by Eq.(\ref{W}). In turn, the value of $q_0$ can be estimated in the isotropic thermodynamic limit  from the sum rule
\begin{equation}
    |\Psi_0|^2 = \int_0^{q_0} W(q) 4 \pi q^2 \ dq,
    \label{Psi0-W}
\end{equation}
which is just an expression of the fact that the intensity of the perfect long-range order $|\Psi_0|^2$ becomes redistributed over small $q$ due to the presence of the topological defects.

Equations (\ref{W}), (\ref{lc}) and (\ref{Psi0-W}) lead to 
\begin{equation}
    l_c = \frac{4 \pi^2 A}{|\Psi_0|^2}.
    \label{lc2}
\end{equation}

In our simulations, we track $W(q)$ after a quench as a function of time, fit the low-$q$ part using Eq.(\ref{W}) and, thereby, obtain the value of the constant $A$ as a function of time. The latter is then substituted to Eq.(\ref{lc2}) to obtain the time dependence of the correlation length.

The resulted time dependencies of $l_c$ are plotted for different quench strengths in Fig.~\ref{quench:correlation_length} and for different lattice sizes in Fig.~\ref{quench:fig:scaling_Lcorr_I2}(b). The comparison of these plots with Figs.~\ref{quench:fig:snapshots_of_defects_18}(c) and \ref{quench:fig:scaling_Lcorr_I2}(a) respectively illustrates, that, as anticipated in Refs.~\cite{zong2019evidence,dolgirev2020amplitude},
the behavior of $I_2$ very closely tracks the behavior of $l_c$.

\subsection{S.VII. Additional snapshots of vorticity after a quench}

\begin{figure*}[htbp]
    \centering
    \begin{tikzpicture}
    \node[inner sep=0pt] (duck) at (0,0)
    {\includegraphics[width=0.32\textwidth]{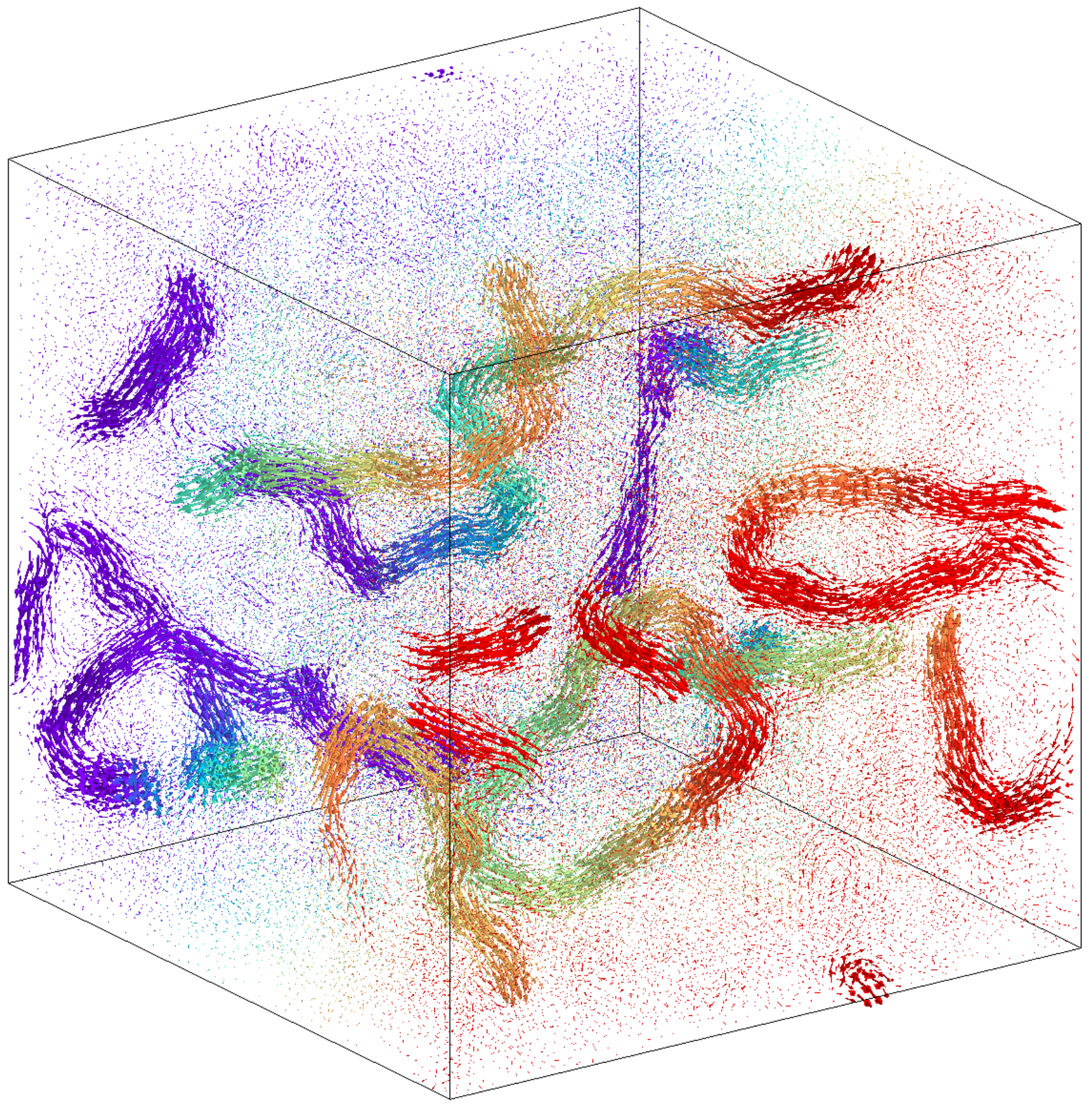}};
    \node[align=center,fill=white] at (-2.4, 2.6) {\textbf{a}};
    \end{tikzpicture}
    ~
    \begin{tikzpicture}
    \node[inner sep=0pt] (duck2) at (0,0)
    {\includegraphics[width=0.32\textwidth]{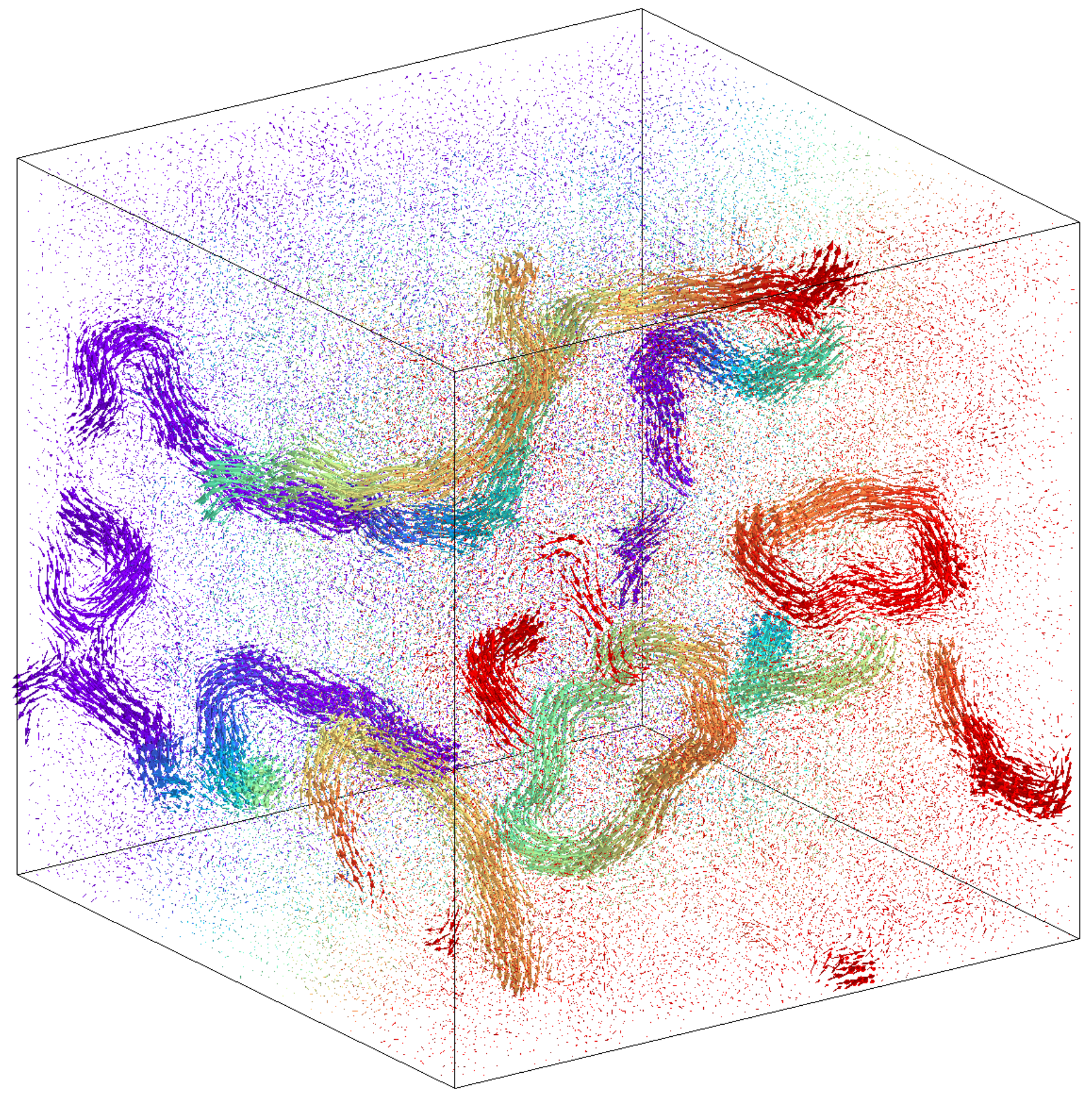}};
    \node[align=center,fill=white] at (-2.4, 2.6) {\textbf{b}};
    \end{tikzpicture}
    ~
    \begin{tikzpicture}
    \node[inner sep=0pt] (duck2) at (0,0)
    {\includegraphics[width=0.32\textwidth]{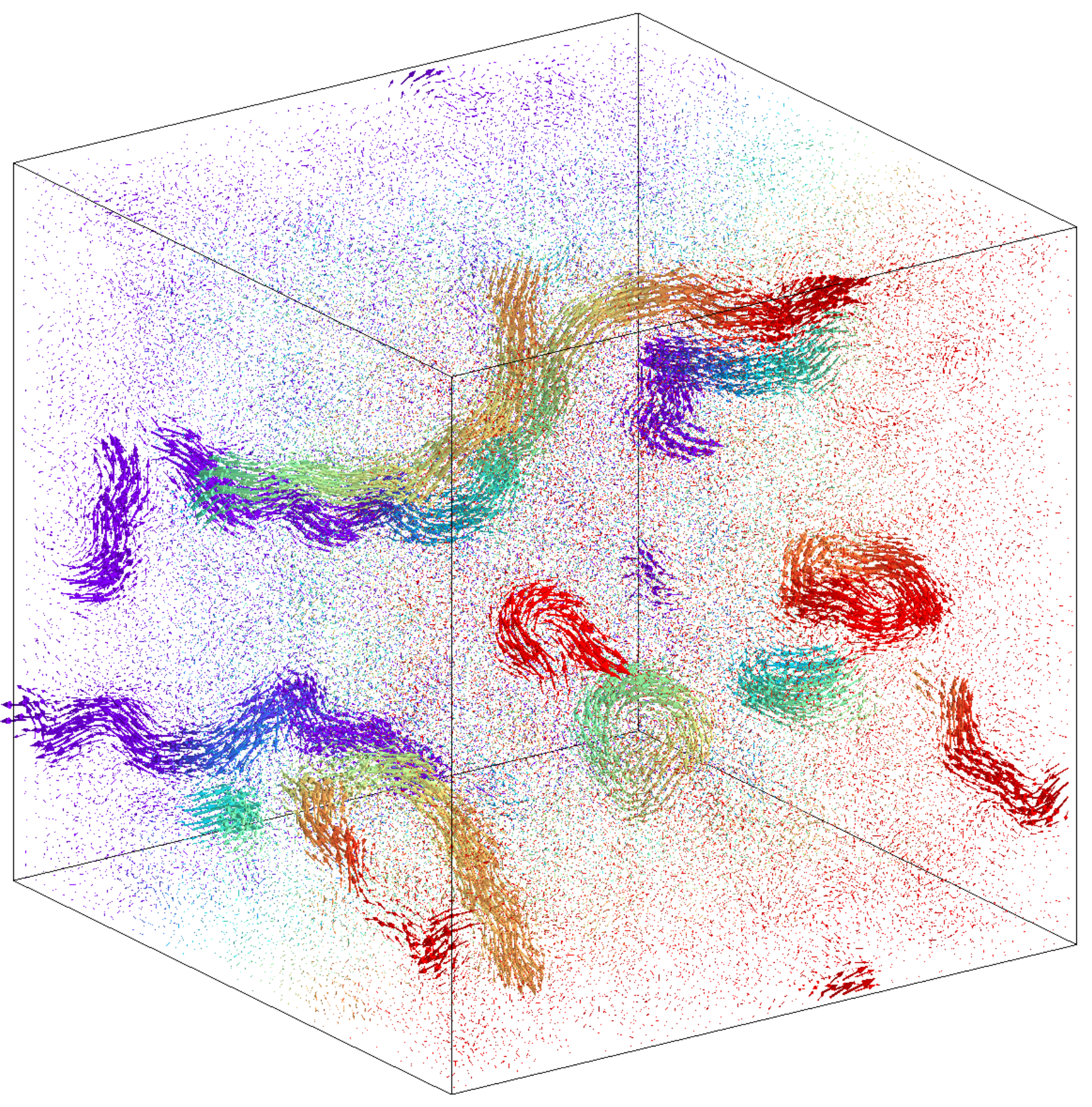}};
    \node[align=center,fill=white] at (-2.4, 2.6) {\textbf{c}};
    \end{tikzpicture}
    \caption{Dynamics of vortices. Snapshots of vorticity for a strong quench from Fig.~\ref{quench:fig:scaling_recovery_time}(b) at times (a)
$t=25 J^{-1}$, (b)
$t=30 J^{-1}$ and (c)
$t=35 J^{-1}$.
\label{quench:fig:snapshots_of_defects_SM}
}
\end{figure*}
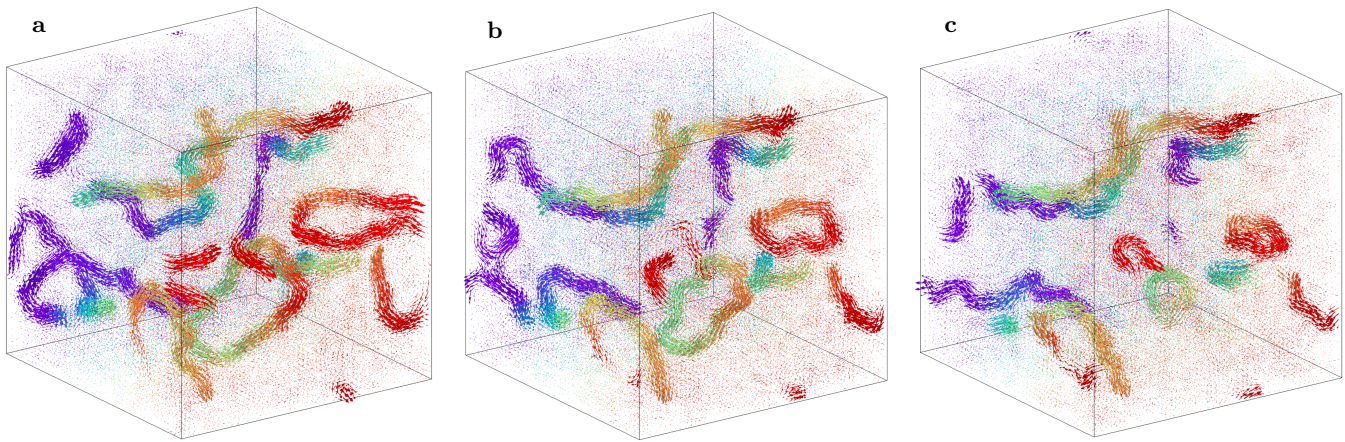

In Fig.~\ref{quench:fig:snapshots_of_defects_SM}, we present three supplemental snapshots of vorticity for a strong quench from Fig.~\ref{quench:fig:scaling_recovery_time}(b) at times $t=25, 30$ and $35 J^{-1}$. The subsequent dynamics of vortices can be observed from these snapshots. The vortices tend to decrease their size, and slowly collapse and disappear.

\subsection{S.VIII. Simulations of anisotropic DGPE lattices}
\label{app::anisotropy}
Many compounds demonstrating $U(1)$ ordering possess pronounced 
quasi-2D anisotropy. For example, 
LaTe$_3$
discussed in the main text may be modelled as a collection of the
Te~planes where the CDW resides, with weak inter-plane coupling.
For such materials it is natural to inquire to what extent the presented 
results are affected by the anisotropy. This question may look
particularly relevant given the incipient  Berezinskii-Kosterlitz-Thouless 
transition. Our simulations, as well as simulations performed in
Ref.~\cite{carlson1999classical},
suggest that, unless the anisotropy is extremely high, its
effects can be accounted for by a simple rescaling of the model 
parameters. 

\begin{figure*}[htbp]
    \centering
    \begin{tikzpicture}
    \node[inner sep=0pt] (duck) at (0,0)
    {\includegraphics[width=0.23\textwidth]{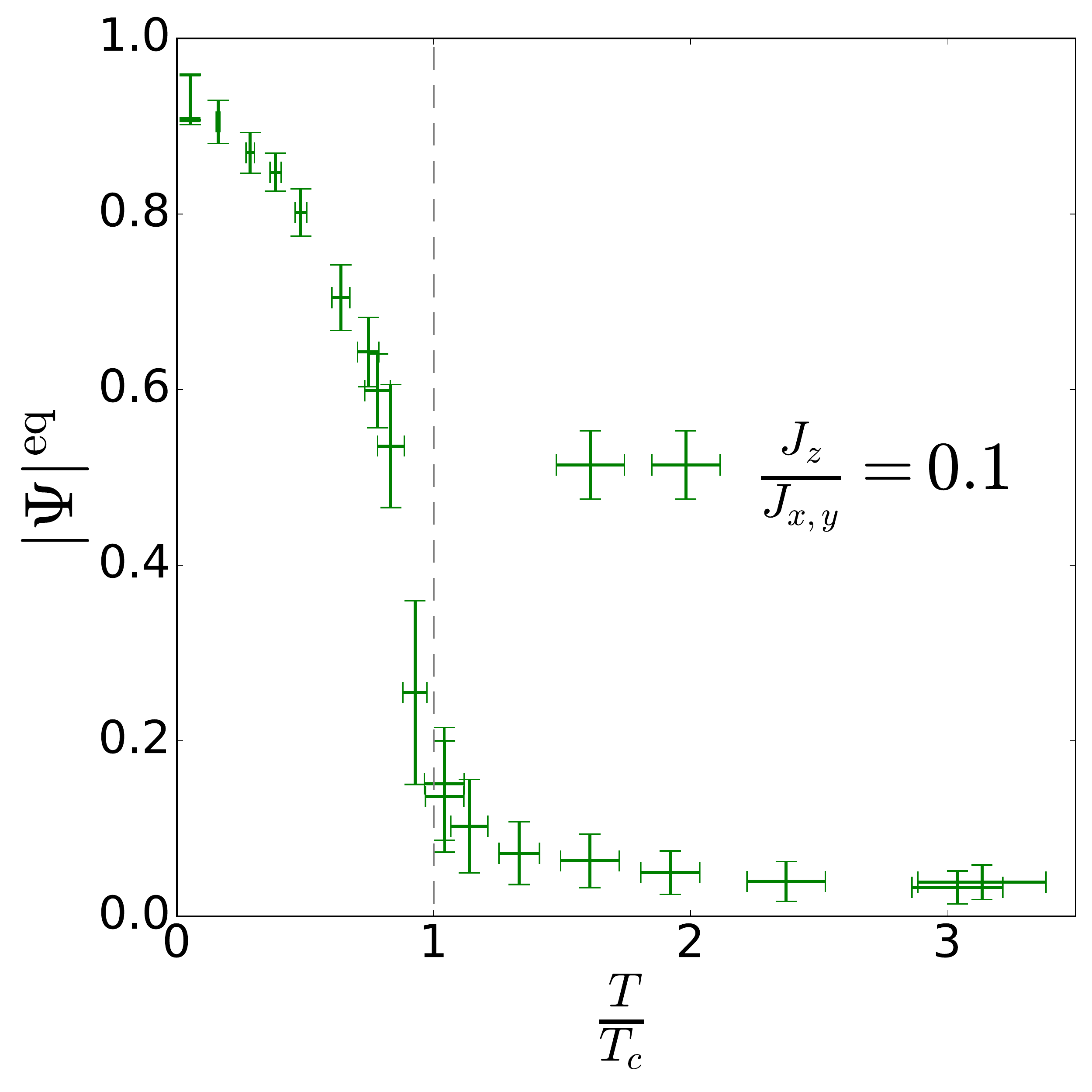}};
    \node[align=center,fill=white] at (0.0, 1.3) {\textbf{a}};
    \end{tikzpicture}
    ~
    \begin{tikzpicture}
    \node[inner sep=0pt] (duck2) at (0,0)
    {\includegraphics[width=0.23\textwidth]{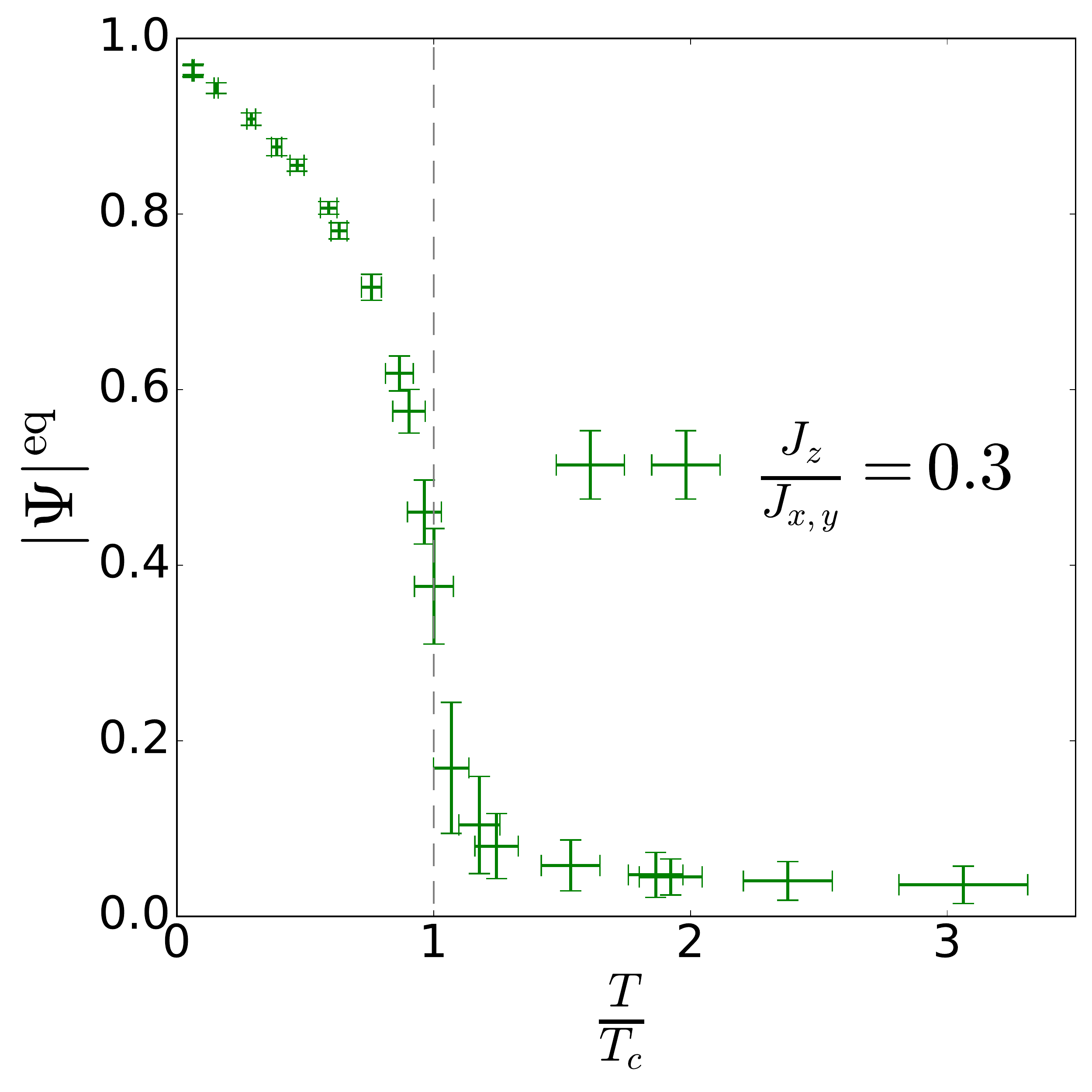}};
    \node[align=center,fill=white] at (0.0, 1.3) {\textbf{b}};
    \end{tikzpicture}
    ~
    \begin{tikzpicture}
    \node[inner sep=0pt] (duck3) at (0,0)
    {\includegraphics[width=0.23\textwidth]{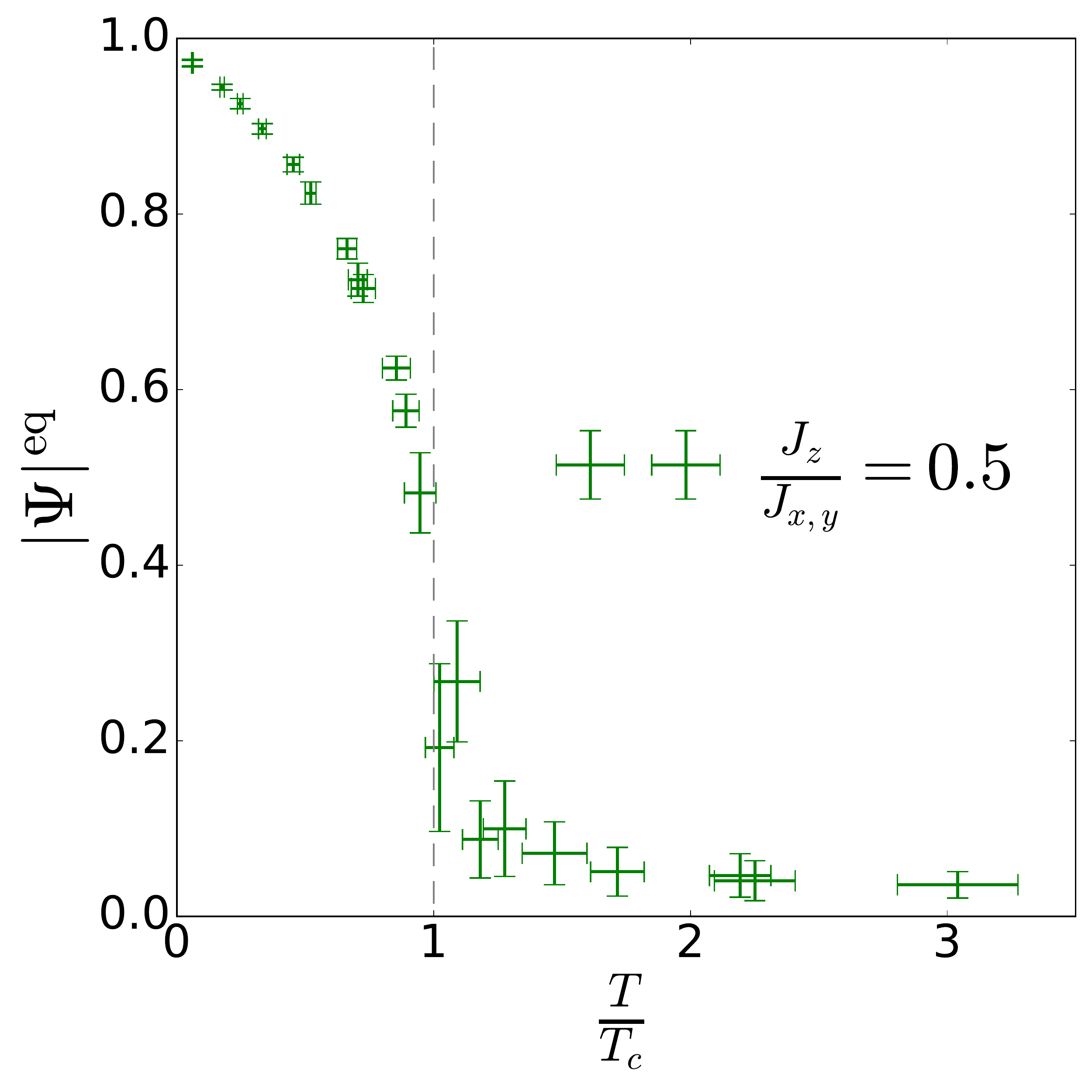}};
    \node[align=center,fill=white] at (0.0, 1.3) {\textbf{c}};
    \end{tikzpicture}
    ~
    \begin{tikzpicture}
    \node[inner sep=0pt] (duck4) at (0,0)
    {\includegraphics[width=0.23\textwidth]{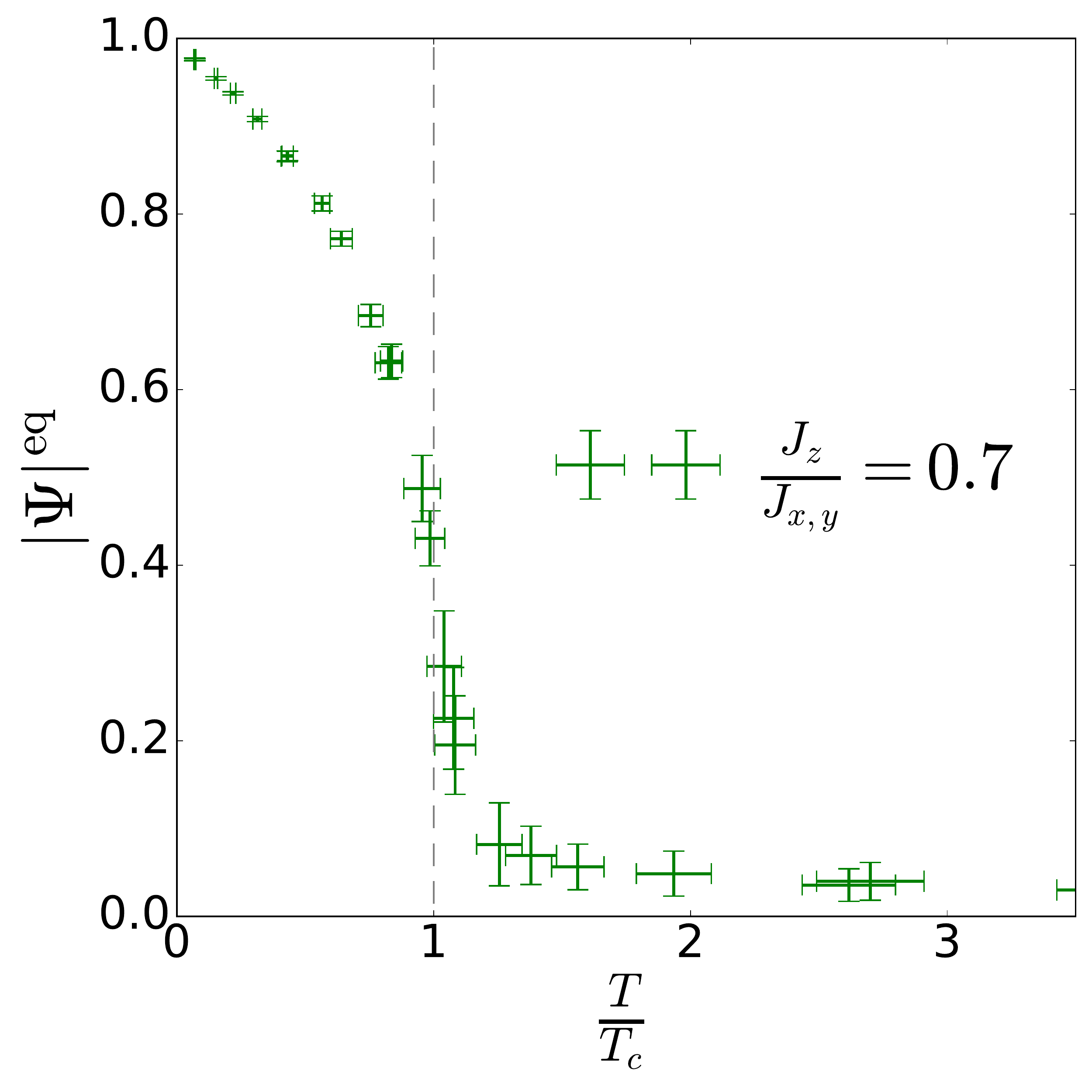}};
    \node[align=center,fill=white] at (0.0, 1.3) {\textbf{d}};
    \end{tikzpicture}
    \caption{The scaling of critical temperature for the DGPE lattices with the following anisotropy parameters: (a) $\frac{J_z}{J_{x,y}} = 0.1$, (b) $\frac{J_z}{J_{x,y}} = 0.3$, (c) $\frac{J_z}{J_{x,y}} = 0.5$, (d) $\frac{J_z}{J_{x,y}} = 0.7$. The critical temperature for each case was estimated within the mean field approximation as $\left(2 + \frac{J_z}{J_{x,y}}\right) / 3 \cdot T_c$, where $T_c$ is the critical temperature of the isotropic DGPE lattice.
\label{quench:fig:anisotropy_on_Tc}
}
\end{figure*}
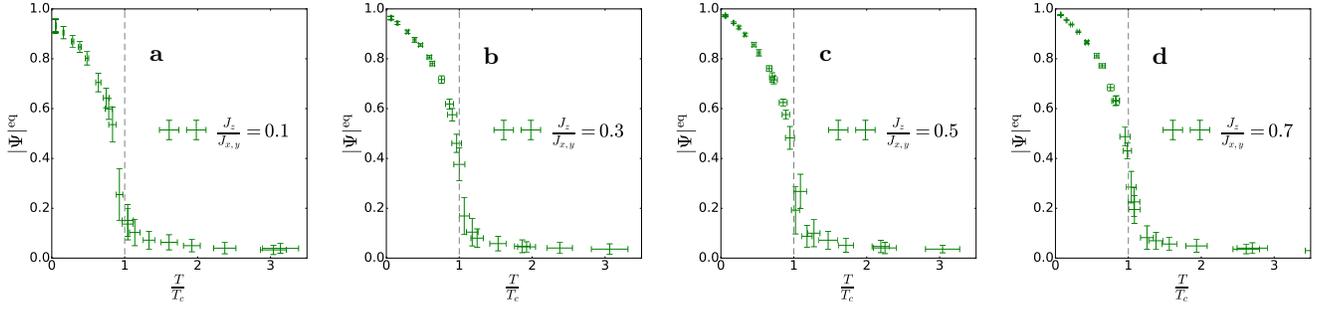

To illustrate this point, we plot the equilibrium value of the order
parameter 
$|\Psi|$
as a function of temperature $T$, for various anisotropies, see
Fig.~\ref{quench:fig:anisotropy_on_Tc}
The four panels in the latter figure are plotted for four different 
values  of the anisotropy ratio
$J_z/J_{x,y}$,
where
$J_{x,y,z}$
represent the coupling constants along the corresponding axes, and
$J_z \leq J_x = J_y$.
We notice that the general dependence
$|\Psi | = |\Psi (T)|$
maintains its overall mean-field-like appearance for all values of
$J_z/J_{x,y}$, with small corrections beginning to appear for $J_z/J_{x,y} = 0.1$. More importantly,
the simple rescaling for the $U(1)$ ordering transition temperature
in the anisotropic model
\begin{eqnarray}
T_c (J_z/J_{x,y}) 
=
\frac{1}{3} \left( 2 + \frac{J_z}{J_{x,y}} \right) T_c (1),
\end{eqnarray}
works well for all anisotropies we had considered. 

As for the Berezinskii-Kosterlitz-Thouless transition, the corresponding
critical point introduces significant new corrections only when the quasi-2D
anisotropy is extremely strong. We believe that our target experimental material
does not enter this regime.

\end{document}